\documentclass[showpacs,aps,prb,reprint,superscriptaddress,twocolumn]{revtex4-1}
\usepackage{amsmath}
\usepackage{graphicx}
\usepackage{subfigure}
\usepackage{psfrag}
\usepackage{color}
\usepackage{rotating}
\usepackage{overpic}
\usepackage{tikz}
\usetikzlibrary{arrows}
\def\GFT{{\bf G}}

\def\br{\mathbf{r}}
\newcommand{\mr}{\mathbf{r}}

\begin{document}

\title{Spontaneous Emission Spectra and Quantum Light-Matter Interactions from a Strongly-Coupled Quantum Dot   Metal-Nanoparticle System}

\author{C. \surname{Van Vlack}}
\email{cvanvlack@physics.queensu.ca}
\affiliation{Queen's University, Dept. of Physics, Kingston Ontario, Canada K7L 3N6}

\author{Philip Tr{\o}st Kristensen}
\affiliation{DTU Fotonik, Technical University of Denmark, Kgs. Lyngby, Denmark}

\author{S. Hughes}
\affiliation{Queen's University, Dept. of Physics, Kingston Ontario, Canada K7L 3N6}



\date{\today}



\begin{abstract}
We investigate the quantum optical properties of a single  photon emitter coupled to  a finite-size metal nanoparticle using a photon Green function technique that  rigorously quantizes the electromagnetic fields. We first obtain pronounced Purcell factors and photonic Lamb shifts for both a 7-nm and 20-nm radius  metal nanoparticle, without adopting a dipole approximation. We then consider a  quantum-dot photon emitter positioned sufficiently near to the metal nanoparticle  so that the strong coupling regime is possible. Accounting for non-dipole interactions, quenching, and photon transport from the dot to the detector, we demonstrate  that the strong coupling regime
should be observable in the far-field spontaneous emission spectrum, even at room temperature.  The  emission spectra show that the usual vacuum Rabi doublet becomes a rich spectral triplet or quartet with two of the four peaks anticrossing, which survives in spite of significant non-radiative decays. We discuss the emitted light spectrum and the effects of quenching for two different dipole polarizations.  
\end{abstract}

\maketitle

\section{Introduction}
The route to photonic vacuum engineering traditionally employs a lossless dielectric cavity system,  exploiting an optical mode with a suitably large quality factor, $Q$, and small effective mode volume, $V$. The local photon density of states (LDOS) scales proportionally with the $Q/V$ factor.
Enhancing the LDOS through the use of small cavities~\cite{Vahala2003} has shown to be a very effective method for increasing the radiative decay rate of an emitter via the Purcell effect~\cite{Purcell-PR-69-681}. In solid state materials, cavities are created using various structural designs, including photonic crystal  lattices
 with defects~\cite{Akahane-NAT-425-944}, and etched micro-pillars made of Bragg reflectors~\cite{qdQED}.
These dielectric cavities have shown some remarkable successes in quantum optics, but the lower limit on \textit{V} in such systems is typically set by diffraction, with {\color{black} $V\approx (\lambda/n)^3$, where $n$ is the refractive index of the cavity.}
Additionally, when one uses quantum dots (QDs), the  narrowband resonance associated with high $Q$  requires very long non-radiative exciton decay times, only achievable at low temperatures.

 In an effort to further increase the LDOS and decrease the system size to sub-wavelength dimensions, it can be advantageous to examine plasmonic systems where light is confined to the surface of a metal and decays evanescently from its surface. For example, a metal nanoparticle (MNP)  supports localized surface plasmons (LSPs)~\cite{Maier-Plamonics} that are tightly confined spatially and not limited by diffraction. This allows coupling between single photon emitters and MNPs\cite{doi:10.1021/nl103906f} which can enhance the LDOS in a system  as small as $10$-$20\,{\rm nm}^3$. 
When the LDOS becomes large enough, it may also be possible to enter the {\em strong coupling} regime where instead of the irreversible process of the emitter
decaying and emitting a photon into the environment (i.e., weak coupling), the emitter can {\em reversibly} exchange the photon with the environment---a process known as vacuum Rabi oscillations. In order for this to happen, the coupling between the emitter and the environment must exceed all possible decay channels. Classical predictions of strong coupling behaviour have been made in the context of metallic dimers~\cite{doi:10.1021/nn100585h}, though it is not known if the splitting survives in the observable spontaneous emission  spectrum. 
This reversible exchange of energy is fundamentally interesting and can 
possibly be useful for applications in coherent control~\cite{PhysRevLett.92.127902}, quantum information processing~\cite{Monroe2002}, and 
lasing/spasing~\cite{PhysRevLett.90.027402,Noginov-Nature-460-1110,Zheludev-NaturePhotonics-2-351}.
With regards to a quantum theory of the light-matter processes in the strong coupling regime, several complications arise in the theoretical description of coupling quantized light to a MNP, including the need to quantize the  fields in a dissipative/lossy medium. Waks and Sridharan~\cite{PhysRevA.82.043845}
 introduced a useful quantized cavity-QED (quantum electrodynamics) treatment of a coupled MNP and a dipole emitter [e.g., a QD], but the MNP was  treated within the dipole approximation~\cite{Carminati2006368}; 
however, it is now well known that 
 the dipole-approximation can yield poor agreement with exact (i.e., non-dipole) calculations obtained from finite-size MNPs---unless placed a few radii from the MNP surface~\cite{Castanie:10,PhysRevLett.96.113002,10.1063/1.443196}. 
{\color{black} Tr\"{u}gler and Hohenester~\cite{PhysRevB.77.115403} have examined the strong coupling dynamics between a molecule and a cigar-shaped MNP employing a mode expansion technique which incorporates the higher order plasmon modes; their quantum approach  assumes a Lorentzian form for the broadening of the modes, via Lindblad superoperators in a master equation formalism~\cite{4683624,Carmichael1999};
this useful non-dipole study predicts the strong coupling regime is possible between a MNP and a molecule though there
is no connection to the emission spectrum.  For dielectric cavity systems, the effects of propagation to a detector is generally assumed to not change the spectral shape of the emitted photons. However for metallic system, because of the losses associated with the MNP and quenching, it is important to compute the generalized light spectrum (i.e., away from the QD position) to first realize if the strong coupling effects are observable, and secondly, to see how the spectral signatures may change and how they would be measured.

In this work, we develop a theoretical formalism that allows one to obtain the emission spectra at any spatial position of the detector.
In Sec.~\ref{sec:theory}, we 
 describe an exact medium-independent quantum optics approach---formulated in terms of  photonic Green functions---to describe the cavity-QED interactions and photon transport between a dipole emitter (QD), a finite-size MNP and
a detector. A schematic of the nanoscale interaction geometry is shown in Fig.~\ref{fig:schematic}(a).
In Sec.~\ref{sec:results}, we present various numerical results
and calculations for the coupled QD-MNP system.
We first calculate the classical Green function above a MNP using  a well established scattering approach   \cite{Tai-GF,L.-W.Li1994}, and
subsequently calculate the  LDOS and  photonic Lamb shift from a nearby dipole emitter, using two different size MNPs (7 nm and 20 nm radius). We find significant enhancements in the LDOS near the MNP surface~\cite{PhysRevLett.96.113002} and simultaneously observe enormous Lamb shifts. We then examine the spectral properties of a QD dipole emitter in the strong coupling regime. 
We compute the far-field spontaneous emission spectrum,
fully accounting for non-Markovian relaxation and propagation effects to the detector.
 The  spontaneous emission spectrum  is shown to yield clear signatures of the strong coupling regime, but is found to be much richer than the usual
vacuum Rabi splitting known from simpler cavity-QED systems (e.g., using dielectric cavities) due to the interplay between  higher-order mode coupling and dipolar-mode coupling; the  non-Markovian spectra yield a spectral triplet or even a quartet of resonances, where two of the peaks anti-cross, thus signalling the strong coupling regime.
We present the strong coupling spectra for two different QD-dipole polarizations and discuss the effects of quenching. 
In Sec.~\ref{sec:discussion}, we give a brief discussion
about possible experimental configurations
for observing our predictions, and in Sec.~\ref{sec:conclusions}
we  conclude.

\begin{figure}[t]
\centering
\subfigure{\begin{overpic}[width=.99\columnwidth]{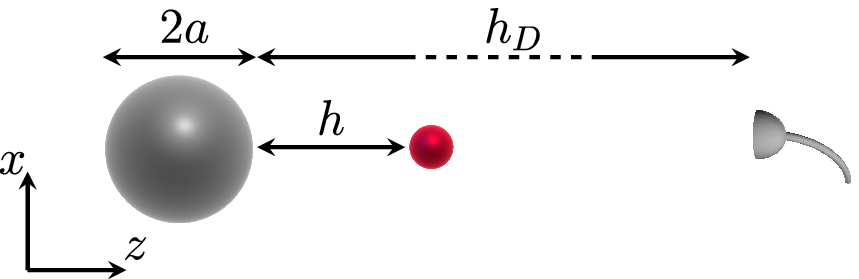}
\put(15,12){\color{yellow}\Large $\boldmath \varepsilon_m$}
\put(27,5){\Large $\varepsilon_b$}
\put(74,0){\large (a)}
\end{overpic}}\\
\subfigure{\hspace{0cm}\begin{overpic}[width=0.99\columnwidth]{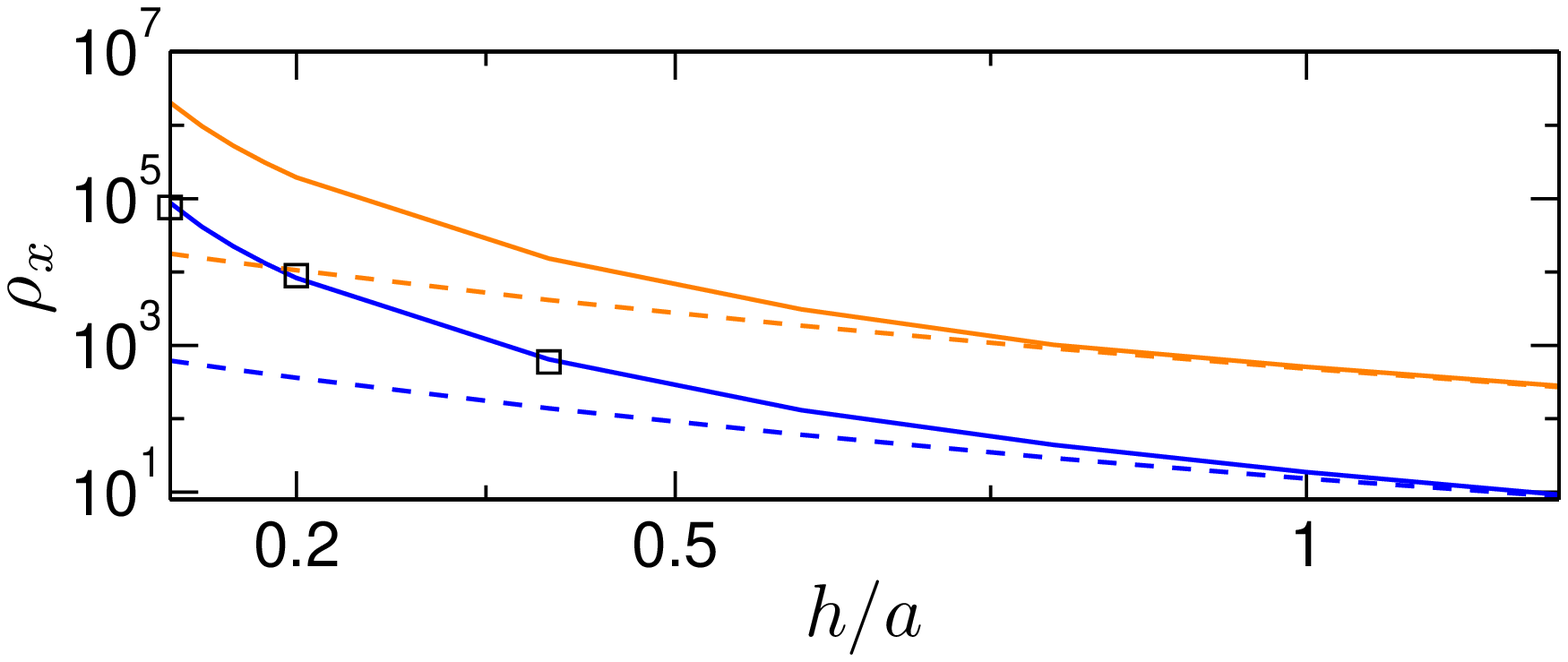}
\put(85,25){\large (b)}
\end{overpic}}
\caption{\label{fig:schematic} (color online) (a) Schematic of the MNP embedded in a background material with permittivity of
$\varepsilon_b$. The  MNP, with radius $a$,
and permittivity, $\varepsilon_m$, is located at the origin. The
single photon emitter (quantum dot) at ${\bf r}_d$ is located at height, $h$, above the surface of the MNP. We also
consider a point-like detector at ${\bf r}_D$ located along the same axis at height $h_D$ 
above the metal surface. (b)  LDOS peak as a function of height above
a 20-nm (blue-dark) and 7-nm (orange-light) spherical silver MNP for an $x$-oriented dipole. The non-dipole result (for the MNP) is given by the solid line and the dipole-approximation result is given by the
dashed line. For comparison, selected finite-difference time-domain results are shown as squares for 20~nm MNPs. }
\end{figure}

\section{Theory}
\label{sec:theory}
\subsection{Green function of a spherical metal-nanoparticle}
The classical 
photon Green function in a medium with 
$\varepsilon({\bf r},\omega)$ (complex dielectric constant) and $\mu=1$, is described through the following equation:
\begin{align}
\nabla\!\times \!\nabla\!\times\! \GFT\! \left(\!\mr,\!\mr'\!;\omega\!\right)
 -  \varepsilon(\!\mr,\omega\!){ k_0^2}\GFT\! \left(\!\mr,\mr';\omega\right)
 = {k_0^2}\delta\!\left(\mr \! - \! \mr'\right) ,
\end{align}
where $k_0 = \omega/c$, where $\omega$ is the angular frequency and $c$ is the speed of light. The dipole-response function (Green function), ${\bf G}$,
can connect to to both classical and quantum light-matter
interactions. For the MNP problem of interest, we will discuss the Green function within and outside the dipole approximation. Typically for small MNPs ($\omega \sqrt{\varepsilon_b} a/c\ll 1$) of permittivity $\varepsilon_m$ embedded in a material with permittivity $\varepsilon_b$, the Green function is obtained through the Dyson equation where we assume that the
spherical MNP response can be modelled through the
metal polarizability function:
\begin{align}
 \alpha_{m}\left(\omega\right) = \frac{{\alpha^0_{m}\left(\omega\right)}}{[{1-\frac{i\alpha_0\omega^3\sqrt{\varepsilon_b}}{6 \pi c^3 a^3}}]},
\end{align}
with the bare polarizability
(i.e., without photon coupling to the environment), \begin{align}
 \alpha^0_{m}\left(\omega\right) = 4\pi \varepsilon_b a^3 \frac{(\varepsilon_m\left(\omega\right) - \varepsilon_b)}{(\varepsilon_m\left(\omega\right) + 2\varepsilon_b )},
\end{align}
 which also accounts for  {\em radiation reaction}~\cite{Draine-AstrophysJ.-333-848}.
Considering the MNP to be located at position $\mr_{ m}$, then the MNP-dipole Green function  is obtained through~\cite{PhysRevLett.105.117701,Novotny-Nano-Optics}:
\begin{align}
\GFT(\mathbf{r},\mathbf{r}') = \GFT_0 (\mathbf{r},\mathbf{r}')
+ \GFT_0  (\mathbf{r},\mr_{m}) \cdot \alpha\, _{m}\GFT_0 (\mr_{m},\mathbf{r}')\,.
\end{align}

To account for the
finite-size nature of the MNP, we also compute the Green function outside the dipole approximation. For these calculations we use
 an established analytical approach where the Green function is expanded in spherical vector functions and the boundary conditions are satisfied at the edge of the sphere~\cite{Tai-GF,L.-W.Li1994}; we relegate the details of this approach to the Appendix.

\subsection{Classical light-matter interactions}
An integral solution for the classical electric field  can be written as 
\begin{align}
 \mathbf{E}\left(\mr , \omega\right ) = \mathbf{E}^0\left(\mr,\omega \right) + \int
\mathrm{d}\mr'\, \GFT \left(\mr,\mr';\omega\right) \cdot \mathbf{P}\left(\mr',\omega\right),
\end{align}

where ${\bf P}$  is a polarization source.
As we will show below, in quantum optics, the
${\bf E}$ and ${\bf P}$ fields  become operators, but ${\bf G}$
remains the same~\cite{Yao-PhysicalReviewB-80-195106,Dung-PRA-57-3931}.
 For a dipole emitter at position ${\bf r}_d$,
 then  $ \mathbf{E}\left(\mr\right) = \mathbf{E}^0\left(\mr\right) +  \GFT \left(\mr,\mr_d\right) \cdot \alpha_d \mathbf{E}\left(\mr_d\right)$,
where the  dipole polarizability of the QD exciton is given by 
 \begin{align}
 \alpha_d(\omega)=\frac{2\omega_d d^2/\hbar\varepsilon_0}{(w_d^2-\omega^2-i\gamma_d\omega)\hbar\varepsilon_0},
 \end{align}
  with
  $\omega_d$ the
 transition frequency, $\gamma_d$  the non-radiative
 broadening of the QD exciton, $d$ the optical dipole moment, $\hbar$ is Planck's constant by $2\pi$ and $\varepsilon_0$ is the permittivity of free space. 
Assuming a QD dipole  of the form $\mathbf{d}=d {\bf n}_i$, then the ({\em projected}) LDOS becomes%
\begin{align}
\label{eq:LDOS}
\rho({\bf r}_d;\omega)= \frac{\mathrm{Im} [\mathbf{n}_i \cdot \GFT \left(\mr_{d},\mr_{d};\omega\right) \cdot \mathbf{n}_i]}{G_0},
\end{align}
where $G_0 =  { k_0^3\sqrt{\varepsilon_b}}/{6\pi}$.
The units of Eq.~(\ref{eq:LDOS})
are 
conveniently  
chosen so that the LDOS is equal to the
Purcell factor~\cite{Purcell-PR-69-681}, which
describes---in a {\em weak coupling regime}---the 
spontaneous emission 
rate, 
\begin{align}
\gamma({\bf r}_d;\omega) = \frac{2d^{2} \rho_{}({\bf r}_d;\omega)G_0}{\hbar\varepsilon_0}.
\end{align}
%
This total electromagnetic (EM)  decay rate includes both radiative and non-radiative coupling with the lossy environment; this modified decay rate obviously depends on the ${\bf G}$ of the medium.
In order to  describe photon propagation  from the
QD to a detector (e.g., to the far
 field), we
also consider the non-local 
propagator, which is defined through the two space-point Green function,
\begin{align}
\rho^{\rm nl}_{ij} \left(\mr,\mr';\omega\right) 
  = \frac{|\mathbf{n}_i \cdot \GFT \left(\mr,\mr';\omega\right)
 \cdot \mathbf{n}_{j}]}{{G_0}}.
\end{align}
The photonic (or {\em anomalous}) Lamb shift is also connected to the
Green function, and is obtained from~\cite{Dung-PRA-68-043816,Yao-PhysicalReviewB-80-195106}
\begin{align}
\Delta \omega\left(\mr_{d};\omega\right) =- \frac{{\mathrm{Re} [\mathbf{d} \cdot \GFT \left(\mr_{d},\mr_{d};\omega\right) \cdot \mathbf{d}]}}{\hbar \varepsilon_0}.
\label{lamb}
\end{align} 
For the  Green function used in Eq.~(\ref{lamb}), i.e., with the same two spatial arguments
Re[${\bf G}({\bf r},{\bf r})$],
 we will neglect the (divergent)  homogeneous-medium contribution  since its effect can be absorbed into the definition of $\omega_d$~\cite{Novotny-Nano-Optics,Vogel-Quantum-Optics}.

The quantities introduced above (e.g., the photon decay rate and the Lamb shift) are well known, and  are perturbative in nature (in their definition). However, this is not a model restriction.
Indeed,  the  theory above can fully include nonperturbative light-matter interactions. 
To reach the {\em strong coupling regime} of cavity-QED, where light-matter interactions
must be included to all orders,
one requires the
dipole-medium coupling rate, $g$, to be larger than any dissipation channels\cite{doi:10.1021/nn100585h,PhysRevB.77.115403}. For a single quasimode of the metal, e.g., ${\bf f}_m({\bf r})$, $g\equiv g_m=\sqrt{\omega_m/2\hbar\varepsilon_0}\,{\bf d}\cdot {\bf f}({\bf r}_d)$ 
so that the vacuum Rabi splitting,  $2g \approx \sqrt{\gamma_{\rm EM}(\rho)\,\gamma_{\rm LSP}/2} \gg \gamma_d,\gamma_{\rm LSP} $. Here $\gamma_{\rm EM}(\rho)$ accounts for all EM decay processes and $\gamma_{\rm LSP}$ is the effective linewidth of the LSP dipole mode; for the purpose of the scaling argument above, we are also tacitly assuming a Lorentzian lineshape for ${\rho(\omega)}$.

 \subsection{Quantum light-matter interactions and the emission spectrum}
 

\begin{figure*}[t]
\vspace{-3cm}
 \centering
 \hspace{-2cm}
 \begin{overpic}[]{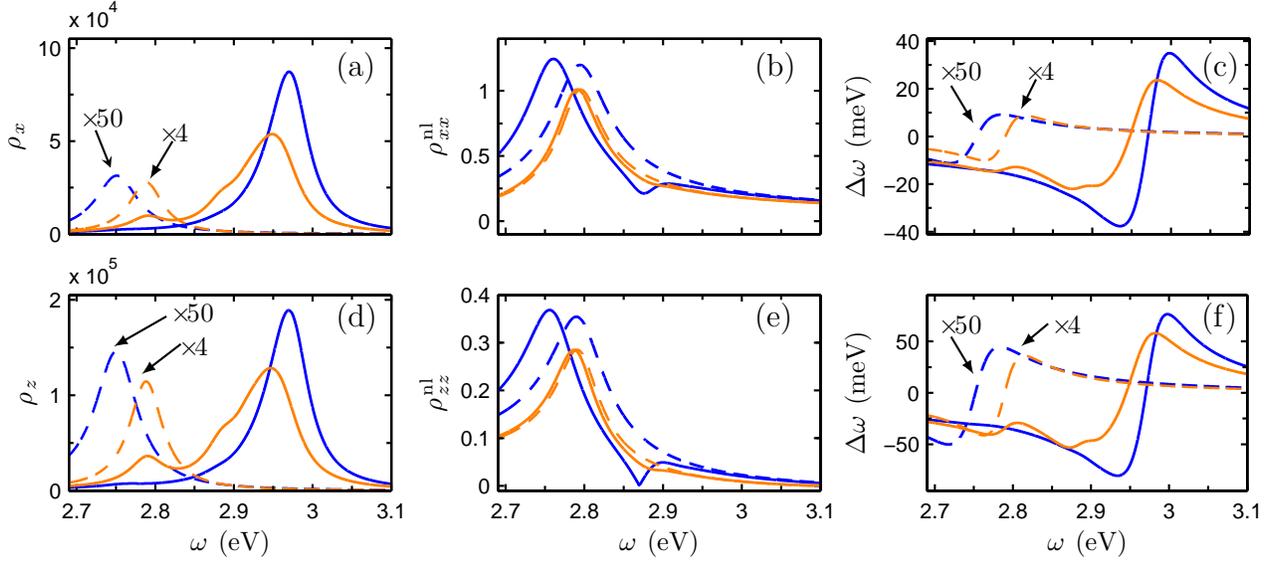}
\put(33,35){\large (a)}
\put(63,35){\large (b)}
\put(95,35){\large (c)}
\put(33,17){\large (d)}
\put(63,17){\large (e)}
\put(95,17){\large (f)}
\put(15.7,28.3){\tikz[>=stealth] \draw[->,thick] (0.1,0.)--(0.3,-0.5);}
\put(19.6,27.5){\tikz[>=latex] \draw[->,thick] (0.0,0.)--(-0.3,-0.5);}
\put(77,31.5){\tikz[>=latex] \draw[->,thick] (0.0,0.)--(0.3,-0.5);}
\put(82,32.5){\tikz[>=latex] \draw[->,thick] (0.0,0.)--(-0.2,-0.4);}
\put(77,13.5){\tikz[>=latex] \draw[->,thick] (0.0,0.)--(0.3,-0.5);}
\put(82,15.5){\tikz[>=latex] \draw[->,thick] (0.0,0.)--(-0.36,-0.3);}
\put(19,13.5){\tikz[>=latex] \draw[->,thick] (0.0,0.)--(-0.52,-0.4);}
\put(17,15.5){\tikz[>=latex] \draw[->,thick] (0.0,0.)--(-0.7,-0.4);}
  \end{overpic}
 \caption{\label{fig:LDOS} (color online) (a) LDOS versus frequency $2$~nm above
a 20-nm (blue-dark) and 7-nm (orange-light) spherical silver MNP. The non-dipole (exact) result is given by the solid line and the dipole-approximation result is given by the
dashed line. 
(b) $\rho_{xx}^{\rm nl}({\bf r}_D,{\bf r}_d)$ versus frequency for $h_d=2~$nm and $h_D=1$~$\mu$m, (c) 
Lamb shift versus frequency. In (a)-(c) we use an $x$-oriented dipole. (d)-(f) as (a)-(c) but for a $z$-oriented dipole. For clarity, dipole results are multiplied by a factor of 50 (20~nm
particle) and 4 (7~nm particle) in graphs (a),(c) and (d),(f).}
\end{figure*}

To  describe the quantum light-matter interaction, we adopt a medium-dependent quantization procedure
for calculating the emission spectrum
from a two-level atom in a  lossy, non-homogeneous environment~\cite{Yao-PhysicalReviewB-80-195106,Dung-PRA-57-3931}. We begin with the complete Hamiltonian of the coupled system,
\begin{eqnarray}
H&=&\!\!\hbar \omega_d \hat{\sigma}^+\hat{\sigma}^-+\hbar\int
d\mathbf{r}\int_0^{\infty}\! \! d\omega_l \,\omega_l
\hat{\mathbf{f}}^\dagger(\mathbf{r},\omega_l)\cdot\hat{\mathbf{f}}(\mathbf{r},\omega_l)\nonumber\\
&-&
[ \hat{\sigma}^+ {\bf d} +  \hat{\sigma}^- {\bf d}] \cdot
\hat{\mathbf{E}}(\mathbf{r}_d) ,
\end{eqnarray}
where $\hat{\sigma}^+,\hat{\sigma}^-$ are the Pauli operators of the QD exciton (located at position $\mathbf{r}_d$), and $\hat{\bf f},\hat{\bf f}^\dagger$ are the bosonic continuum field creation/annihilation operators of the total electric field operator (including interactions with the QD), which are indexed in the Hamiltonian with continuous eigenfrequencies $\omega_l$. The electric field operator is related to the bosonic field operators through~\cite{Dung-PRA-68-043816},
\begin{multline}
\hat{\mathbf{E}}(\mathbf{r},t) =  \hat{\mathbf{E}}^0(\mathbf{r},t) +
i\sqrt{\frac{\hbar}{\pi\varepsilon_0}}\int_0^{\infty}\! d\omega_l\int
d\mathbf{r}'\mathbf{G}(\mathbf r,\mathbf r';\omega_l)\\
\cdot\sqrt{\varepsilon_I(\mathbf{r}',\omega_l)}\,\hat{\mathbf{f}}(\mathbf{r}',\omega_l;t)+ {\rm H.c.},
\end{multline}
where $\varepsilon_I$ is the imaginary part of the permittivity  and $\hat{\mathbf{E}}^{0} (\mathbf{r},t)$ is the  {\em free field}, i.e., the field that exists without the
presence of the QD.
To proceed we will adopt the weak excitation approximation, so that we assume at most one quantum in the system (this approximation is exact when the initial field is in vacuum). Using the Heisenberg equations of motion, and Laplace transforming to the spectral domain, we can subsequently obtain explicit expressions
for $\hat \sigma^{+},\hat \sigma^-$ and $\hat {\bf f},\hat {\bf f}^\dagger$. %
The total electric field operator is then~\cite{Yao-PhysicalReviewB-80-195106}:
\begin{align}
\label{eq:E_tot}
\hat{\mathbf{E}}\left(\br,\omega\right) &=  \hat{\mathbf{E}}^0(\mathbf{r},\omega) \nonumber \\ 
 &+ \int   \frac{{\rm Im}\mathbf{G} \left(\br,\br_d;\omega_l\right) \cdot \mathbf{d}}{\pi \varepsilon_0}  \frac{\hat{\sigma}^-(\omega)+\hat{\sigma}^+(\omega)}{\omega - \omega_l} , \nonumber \\
&= \hat{\mathbf{E}}^0(\mathbf{r},t) +\frac{1}{\varepsilon_0}\mathbf{G} \left(\br,\br_d;\omega\right) \cdot \mathbf{d} [\hat{\sigma}^-(\omega)+\hat{\sigma}^+(\omega)] ,
\end{align}
in which we have used the relation $\frac{i}{\omega_l - \omega +i\epsilon_+} = \pi \delta \left(\omega_l - \omega\right) + i{\rm P}\left(\frac{1}{\omega_l-\omega}\right)$ has been used, with  ${\rm P}$ is the principle value.
The dipole operators are given by,%
\begin{align}
& \hat{\sigma}^-(\omega)+\hat{\sigma}^+(\omega) = \nonumber \\
& \ \ \ \frac{-i\left[\hat{\sigma}^-(t=0)(\omega+\!\omega_d)
+\hat{\sigma}^+\!(t\!=\!0)(\omega-\omega_d)\right]}
{\omega_d^2-\omega^2-2\omega_d\,{\bf d}\!\cdot\!\mathbf{G}(\mathbf{r}_d,\mathbf{r}_d;\omega)\cdot{\bf d}/\hbar\varepsilon_0}.
\end{align}

The light spectrum is defined through $S({\bf r},\omega)=\int_0^\infty dt_1\int_0^\infty dt_2 e^{i\omega (t_2-t_1)} \langle [\hat{\mathbf{E}}\left(\br,t_1\right)]^\dag\hat{\mathbf{E}}\left(\br,t_2 \right)\rangle$, which gives $S({\bf r},\omega)=\langle [\hat{\mathbf{E}}\left(\br,\omega\right)]^\dag\hat{\mathbf{E}}\left(\br,\omega\right)\rangle$. For $\br = \br_D$, and assuming
an initially-excited QD  exciton in vacuum, one obtains~\cite{Yao-PhysicalReviewB-80-195106} the emitted light-spectrum, analytically, 

\begin{align}
&S({\bf r}_D,\omega) = \nonumber \\ 
& \ \ \ \left|\frac{\mathbf{d} \cdot\GFT\left({\bf r}_D,{\bf r}_d;\omega\right) \left(\omega + \omega_d\right)/\varepsilon_0}{\omega_d^2 - \omega^2 -i\omega \gamma_d-2\omega_d\, \mathbf{d} \cdot \GFT\left({\bf r}_d,{\bf r}_d;\omega\right)\cdot \mathbf{d}/\hbar \varepsilon_0 } \right|^2,
\label{eq:spectrum}
\end{align}
where the point detector is assumed to be at position $\mathbf{r}_D$ above the center of the MNP. We highlight that this final spectrum is  exact in both weak and strong coupling limits.
In order to more clearly extract the physics associated with propagation and quenching, we will also examine the dipole or polarization spectrum: 
%
\begin{align}
 P(\omega)& \equiv \langle \hat{\sigma}^+(\omega)\hat{\sigma}^-(\omega)\rangle \nonumber \\
&= \left|\!\frac{1}{\omega_d^2 - \omega^2\! -\!i\omega \gamma_d-2\omega_d \mathbf{d}\, \cdot\! \GFT\left({\bf r}_d,{\bf r}_d;\omega\right)\!\cdot\! \mathbf{d}/\hbar \varepsilon_0 } \right|^2,
\label{eq:spectrum_dot}
\end{align}
which contains important information about the local  dot dynamics. Worth  to note is that Eqs.~(\ref{eq:spectrum}) and (\ref{eq:spectrum_dot}) are applicable in {any} lossy, non-magnetic inhomogeneous system, provided it is possible to calculate the Green function, which illustrates the strength of our technique. We also remark that it is relatively straightforward to include multiple QDs within this formalism~\cite{PhysRevB.83.075305}.

Before closing this theory section, we make a few general comments on the
 form of the QD  non-radiative decay rate, $\gamma_d$.
 This broadening mechanism is likely caused by electron-phonon scattering and {\em pure dephasing}, which is especially important  at elevated temperatures.
Although we have not distinguished the mechanism of pure dephasing from 
an effective decay rate in the polarizability, the computed
spectrum maintains precisely the same spectral shape for our
chosen initial conditions~\cite{PhysRevB.83.165313}; so the distinction of pure dephasing is not necessary for computing the vacuum  spectrum.
However, if one knows the precise spectral form of the QD polarizability, including
the influence of electron-phonon scattering, then only a small modification is needed
in the above formulas~\cite{PhysRevB.83.165313}. 
For the calculations that follow below, we will
adopt broadening values similar to colloidal dots
at room temperature~\cite{PhysRevB.82.165435},
with $\gamma_d=10$-$20~$meV. 
Note also, that
since 
the dominant decay is from non-radiative coupling to the lossy MNP,
 the details of the bare exciton decay are less important here (e.g., in comparison to coupling to a dielectric cavity system).
An alternative quantum optics approach can include
phonon interactions at the level of a polaron master equation~\cite{PhysRevLett.106.247403,PhysRevB.83.165313}.

\section{Results}
\label{sec:results}
\subsection{Weak coupling regime: Purcell factors and Lamb shifts}
For our numerical calculations,
we assume a MNP with a permittivity given by the Drude model,
$\varepsilon_m = \varepsilon_{\infty} - {\omega_m^2}/({\omega^2 - i \gamma_m \omega)},$
and take the parameters  typical for silver~\cite{doi:10.1021/nn100585h}: $\varepsilon_{\infty} = 6$, 
$\omega_m = 7.90$~eV and 
$\gamma_m = 51$~meV; this gives an estimated $\gamma_{LSP} = 60$~meV and $\gamma_{LSP} = 75$~meV for 7~nm and 20~nm particles, respectively, in the regime where the dipole approximation is valid.
We consider a dipole emitter located 
$h=2$~nm above a 7-nm and a 20-nm MNP. For the single photon emitter (QD exciton), we consider both $x$-oriented and $z$-oriented dipoles with a 
dipole moment of $d=24\,$Debye ($\approx 0.5~e\,{\rm nm}$) which is comparable to (or less) than the dipole moment used in other works that model QDs 
coupled to metals~\cite{PhysRevB.77.115403,PhysRevB.77.115403,doi:10.1021/nn100585h}.

In Fig.~\ref{fig:schematic}} (b) we show the LDOS versus height using both the non-dipole  and  dipole calculations. 
We observe convergence between the solutions with the analytic and the dipole-approximation
only for $h>2a$, in agreement with  Ref.~\onlinecite{Castanie:10}.
Additionally for the 20-nm radius MNP, we plot the same calculations performed using finite-difference time-domain calculations~\cite{lumerical} (squares) and a $1$-nm grid size (finite-size emitter);  we observe excellent agreement between these two different methods.

  \begin{figure}[t!]
\includegraphics[width=0.99\columnwidth]{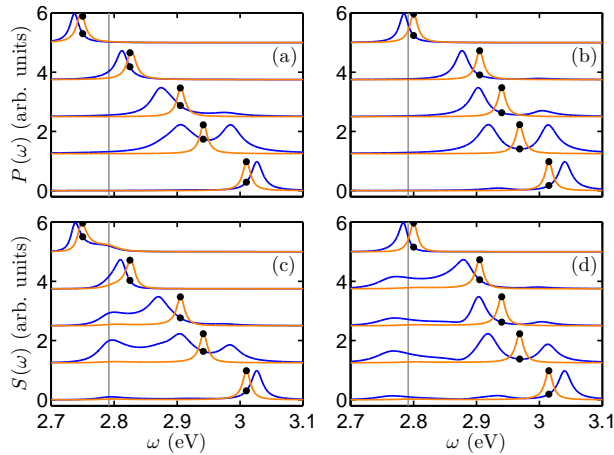}
 \caption{\label{fig:anti} (color online) (a)+(c) 7-nm particle [(b)+(d) 20-nm particle] with an emitter 2~nm
from the surface. Graphs (a)-(b) and (c)-(d) show the normalized effective
particle and far-field spontaneous emission spectra, respectively, using the non-dipole  result (blue-dark line) and the dipole-approximation (orange-light line) for an $x$-oriented dipole.
Transition frequencies are indicated
by black dots on the curves.
The thin grey line in all graphs indicates the LSP resonance (at the maximum of $\alpha_m$).}
\end{figure}

In Figs.~\ref{fig:LDOS}(a) and \ref{fig:LDOS}(d) we plot the LDOS as a function of frequency for $h=2$~nm above the 7-nm and 20-nm MNPs for $x$-oriented and $z$-oriented dipoles, respectively. We immediately notice that the LDOS peaks are far separated in energy when compared to the dipole result, which is caused
by the essential contribution from higher-order modes~\cite{10.1063/1.3532101}. We also see that the LDOS peak for both nm-size particles is comparable, but
the LDOS peak is slightly shifted between the two different sized particles. When comparing between $x$-oriented and $z$-oriented dipoles we see that the LDOS is larger for the latter case by about a factor of 2. 
Also note that the difference in the dipole-approximation for the 7-nm particle compared to the 20-nm particle is mainly due to the fact that they have different center to center distances; the shorter distance gives a larger result 
because of the scaling of the free space Green function 
in the near field, i.e., ${\bf G}_{\rm free}({\bf r}_d,{\bf r}_m) \propto |{\bf r}_d-{\bf r}_m|^{-3}$.

We next consider the non-local propagator in Figs.~\ref{fig:LDOS} (b) and (e) for $x$-oriented and $z$-oriented, respectively; this propagator  is needed to account for light propagation from the dipole emitter to the detector. 
 The detector is assumed to be at a height of $1~\mu$m above the MNP surface. For the 20-nm particle, the non-dipole calculations for   $\rho^{\rm nl}$  is spectrally peaked near $\omega\approx 2.76$~eV, which does not coincide with the peak of the LDOS ($\approx2.97$~eV); however,  the peak in the  $\rho^{\rm nl}$ using a dipole-approximation  is shifted to 2.79 eV. 
We also observe an additional peak located near 2.9~eV, and we show below how
this complex lineshape affects the  spontaneous emission spectrum.
We can contrast these 20-nm MNP findings with the 7~nm results, where the $\rho^{\rm nl}$  in the dipole-approximation agrees quite well with the exact result---although we begin to observe a small shoulder in the non-dipole result which indicates a second peak.
Both non-local propagator peaks in this region are located at
2.8~eV, which is the same location as the dipole peaks seen in the LDOS and again the difference between $x$-oriented and $z$-oriented dipoles is about a factor of 2; but now the $xx$-component of the non-local propagator is the larger (suggesting less
quenching).

 \begin{figure}[t]
\includegraphics[width=0.99\columnwidth]{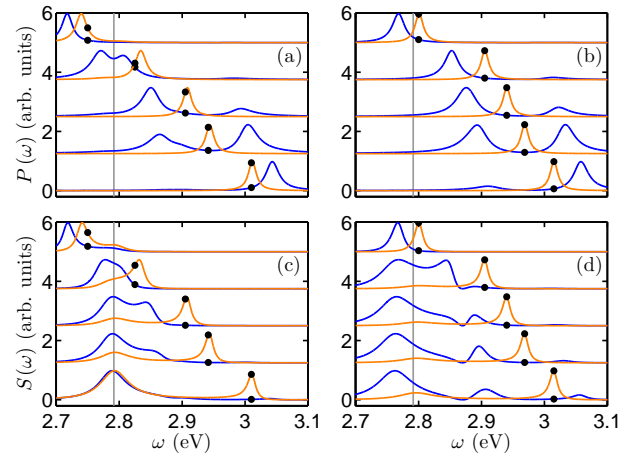}
 \caption{\label{fig:anti_r} (color online) As in Fig.~(3), but with a $z$-oriented QD dipole. }
\end{figure}

\begin{figure*}[t]
\begin{overpic}[]{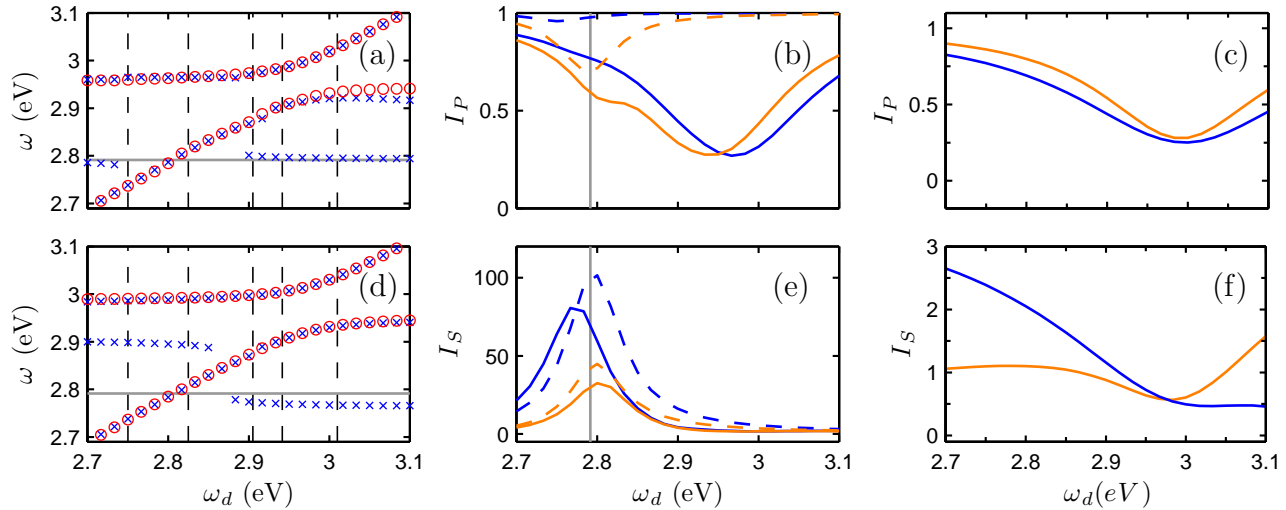}
\put(28,35){\large (a)}
\put(60,35){\large (b)}
\put(94,35){\large (c)}
\put(28,17){\large (d)}
\put(60,17){\large (e)}
\put(94,17){\large (f)}
  \end{overpic}
 \caption{\label{fig:anti2} (color online) (a) 7-nm particle [(d) 20-nm particle] with an emitter 2~nm
from the surface tracking the  particle spectral peaks (red
circles) and light emission spectral peaks (blue crosses) as a
function of QD  frequency for an $x$-oriented QD.  The dashed
lines correspond to the location of the QD transition frequencies used
in the graphs in Fig.~\ref{fig:anti}. (b) Integrated particle spectra for an $x$-oriented QD normalized by the integrated particle spectrum without the MNP (i.e., free space) as a function of QD transition frequency for 20~nm MNP (blue-dark line) and 7~nm MNP (orange-light line) for dipole (dashed line) and non-dipole (solid line) calculations. (c) Integrated particle spectrum $2$~nm above a metallic half space normalized by integrated particle spectrum in free space as a function of QD transition frequency for $z$-oriented QD (blue-dark line) and $x$-oriented QD (orange-light line). (e) as (b) but with the integrated far field spectrum normalized by the integrated far field spectrum without the MNP. (f) as (c) but with the integrated far field spectrum normalized by the integrated far field spectrum in free space. The thin grey lines in (a),(b),(d),(e) indicate the LSP resonance.}
\end{figure*}

Figures~\ref{fig:LDOS}~(c) and (f) show the photonic Lamb shift for $x$-oriented and $z$-oriented dipoles, respectively, for both MNP sizes. Again the ({invalid}) dipole solutions are plotted for reference. The Lamb shifts at this height are quite large, giving a maximum  frequency shift of $|\Delta \omega|_{\rm max}/\omega = 7.9\times 10^{-3}$ for the 7-nm particle and $|\Delta \omega|_{\rm max}/\omega = 1.28\times 10^{-2}$ for the 20-nm particle in the $x$ direction, and a maximum frequency shift of $|\Delta\omega|_{\rm max}/\omega = 2.02\times 10^{-2}$ for the 7-nm particle and $|\Delta \omega|_{\rm max}/\omega = 2.75\times 10^{-2}$ for the 20-nm particle in the $z$ direction. 
For comparison, at $\omega\sim 3~$eV, an exciton linewidth of $\gamma_d\sim 15\,$meV corresponds to $\gamma_d/\omega \approx 5\times 10^{-3}$, so the largest frequency shift in Fig.~\ref{fig:LDOS}~(f) is more than 5 times the exciton linewidth (even at room temperature), which, to our knowledge, is much larger than any previously reported result. For  photonic crystal systems\cite{Wang-PRL-93-073901},  $|\Delta \omega|_{\rm max}/\omega \approx 4\times 10^{-5}$ has been reported, and for negative index metamaterial slabs~\cite{Yao-PhysicalReviewB-80-195106},   $|\Delta \omega|_{\rm max}/\omega \approx 5\times 10^{-4}$ has been predicted.

\subsection{Strong coupling regime and  emitted spectrum}
Motivated by the significant enhancements seen in Fig.~\ref{fig:LDOS}, we next study the nonperturbative strong-coupling regime, and calculate  both the particle (or polarization) spectrum and the spontaneous emission spectrum of the field (Figs.~\ref{fig:anti} and \ref{fig:anti_r}). As mentioned above, for all calculations we use $\gamma_d=15$~meV which corresponds to the decay of a typical QD exciton at room temperature~\cite{PhysRevB.82.165435}. Such a large decay would completely dominate semiconductor cavity systems, where the best (maximum) vacuum Rabi splittings are around 0.1~meV~\cite{qdQED}. For the 7~nm particle, with an $x$-oriented QD, the non-dipole  result (i.e., not treating the MNP as a dipole) for $P\left(\omega\right)$ [Fig.~\ref{fig:anti}~(a)] shows that there is a clear anticrossing, and  the spectral location of strong coupling is evidently not located at the dipole LSP of 2.7915~eV;  rather, it is much higher in energy at 2.9415~eV corresponding to the peak of the LDOS ($\sim2.9489$~eV). In contrast, the dipole result shows no indication of strong coupling.  When looking at the far field spectrum [see Fig.~\ref{fig:anti}~(c)], it is more difficult to observe an anticrossing as the weighting provided by $\rho^{nl}$  causes the peaks to broaden and become  more asymmetric. Additionally, there is a clear peak at 2.7885~eV which corresponds to the peak in $\rho^{\rm nl}$; this spectral peak  is not observable in the particle spectra; this additional peak also shows up when using the dipole approximation but we emphasize that it is only due to photon propagation from the MNP/emitter system to the detector and is extremely small. We highlight that these additional spectral features in the emission spectrum
are quite different to a dielectric cavity system.

From the calculations above, it is also clear that the predicted Lamb shifts are observable in the spectra as the exciton spectral peaks in both the particle and the far-field spectra are substantially shifted in energy. Figures~\ref{fig:anti}(b) and ~\ref{fig:anti}(d) show the particle and light spectra for the 20-nm MNP, and we observe many similar features to the 7-nm spectra; however the splitting between the peaks in the particle spectrum is notably larger for the 20~nm particle compared to the 7~nm particle. In the far field spectra, we also see that there are significant qualitative differences as the QD frequency is tuned; however the anticrossing region is observed at similar QD detunings.
 This shows that the strong coupling is clearly an observable effect, even with metal losses.
 
Figure \ref{fig:anti_r}  shows similar results to those shown in  Fig.~\ref{fig:anti}, but with  a $z$-oriented exciton (dipole). The larger LDOS  for this polarization manifests in an increased Rabi splitting in both particle and far field spectra, however the peaks at higher energies are more difficult to observe on this scale due to the  smaller value of the non-local propagator in the higher frequency range [see Fig.~\ref{fig:LDOS}(e)].



To further examine the anticrossing behavior of strong coupling, we have located the maxima in the particle spectra (red circles) and far field  spectra (blue crosses) for various QD transition frequencies, and show these in Fig.~\ref{fig:anti2}~(a) for 7-nm particles, and Fig.~\ref{fig:anti2}~(d) for 20-nm particles using an $x$-oriented QD (similar results are found for a $z$-oriented dipole). In both the particle and light emission spectra, a clear anticrossing is observable indicating a vacuum Rabi splitting of around $2g\approx 79$~meV for 7-nm particles and $2g\approx 95$~meV for 20-nm particles; note that $2g> 120$~meV for both sizes of MNP for the $z$-oriented QD (not shown). Clearly this observation does not correspond to the location of the lowest-order dipole mode (indicated by the thin grey lines in Fig.~\ref{fig:anti}) but is due to the coupling to higher order plasmon modes. For the emitted light spectrum, the effects of propagation add additional peaks, however the vacuum Rabi splitting is well maintained even with non-radiative quenching. This finding is not at all clear unless one properly accounts for propagation to the detector. 

In Figs.~\ref{fig:anti2}(a) and (d) we discern an additional peak in the far field at the location of the dipole mode which is due to light propagation (via  $\rho^{\rm nl}$), and in Fig.~\ref{fig:anti2} (d) there is a fourth peak which is due to the dip located in $\rho^{\rm nl}$  that only occurs for the 20-nm particle, but becomes too small to resolve after $\omega_d\approx2.85$~eV.
These spontaneous emission spectra contain highly non-Lorentzian lineshapes as well as essential non-dipolar interaction effects. Furthermore, any predictions of strong coupling with MNPs {must} include higher order mode coupling as they will dominate the dynamics before it is ever possible to achieve strong coupling using the dipole mode (at least for our chosen parameters).

Finally, we study the quenching effects in more detail. In 
Figs.~\ref{fig:anti2}(b) and \ref{fig:anti2}(e) we calculate the integration of the particle/far field spectrum as a function of $\omega_d$, and we normalize this to the integrated free-space value. We define the following
 integrated spectral quantities, 
$I_P(\omega_d)$ (integrated particle spectrum ) or $I_S(\omega_d)$ (integrated far field spectrum), which are computed as follows: 
\begin{align}
I_{P}(\omega_d)&=\frac{\int_0^\infty P(\omega,\omega_d)\, d\omega}{\int_0^\infty P_{\rm hom}(\omega,\omega_d)\, d\omega }, \\
I_{S}(\omega_d)&=\frac{\int_0^\infty S(\omega,\omega_d)\,d\omega}{\int_0^\infty S_{\rm hom}(\omega,\omega_d)\, d\omega }.
\end{align}
These integrals give 
the likelihood of detecting a photon emitted by a QD, and
the values in the vicinity of the MNP are normalized to the values
that would be obtained from a  QD in vacuum (for this particular particle/detector geometry). We show  MNP dipole-approximation (dashed) and non-dipole (solid) results for a 7-nm MNP (orange-light line) and a 20-nm MNP (blue-dark line). For the integrated particle spectra, $I_P$, we see that, in terms of emitted flux, quenching is much more problematic for the non-dipolar result compared to the dipolar result. The region of greatest quenching is where the LDOS is peaked giving a maximum reduction to $I_P\approx 0.3$ in the region of the anti-crossing. Such an observation would lead one to believe that MNPs appear to absorb the majority of the emitted photons. However, in the far-field spectrum, $I_S$, we see a dramatic {\em increase}  in the relative number of photons detected for QDs located near the LSP of 30 (80) for 7~nm (20~nm). Even in the anticrossing region, the enhancement is $\approx 2-3$ compared to a QD in free space.  This enhancement in the integrated far field spectrum shows that even in the frequency region where photons appear to be dominated by non-radiative effects, the MNP compensates by acting as an antenna making the detection of far field radiation more efficient. 

To help further clarify the physics of metallic quenching, we also compare the MNP case with a metallic half space, where we calculate the Green function using a well known multilayer technique~\cite{Paulus-PRE-62-5797,Yao-PhysicalReviewB-80-195106}. We initially verify in Fig.~\ref{fig:anti2}(c) that a simple metallic half space suffers similar quenching to the MNP in the particle spectrum; however, it feels much more quenching in the far field spectrum [Fig.~\ref{fig:anti2}(f)] with $I_S$ lower by about two orders of magnitude for both $z$-oriented (blue-dark line) and $x$-oriented (orange-light line) QDs.  It is worth noting that the reduction of $\gamma_d$ to values typical for QDs at cryogenic temperatures ($\approx \mu$eV) results in much more dramatic quenching in both the particle spectrum and the far field spectrum for QDs coupled to MNPs; for example, using  $\gamma_d<50\,\mu$eV results in  $I_S<1$ over the entire frequency range showing that the antenna effect of the MNP is unable to overcome the quenching in the case of sharp QD linewidths.

For these quantum optical studies above, we have deliberately chosen a rather large dipole moment ($d=24\,$ Debye) to enable the strong coupling regime.
For smaller dipole sizes, e.g., with  $d=12$~Debye for the $x$-oriented dipole, or $d=8$~Debye for a $z$-oriented dipole, we obtain qualitatively similar
strong coupling results but with smaller vacuum Rabi splittings.  There is also the potential to see strong coupling with even lower QD dipole moments, if one uses  MNPs with non-spherical shapes, e.g. cigar shapes~\cite{PhysRevB.77.115403}.
For much lower dipole moments then the strong coupling effect of course vanishes, although dimer~\cite{doi:10.1021/nn100585h,Koenderink:10} configurations may help to increase the LDOS to a sufficiently larger value.
\section{Discussion}
\label{sec:discussion}
We briefly discuss some potential experimental configuration for observing the effects presented above. There are several possible experimental scenarios that are likely within reach of current nano-fabrication techniques~\cite{Benson2011}. One example could  involve spin coating colloidal QDs onto a substrate, locating the dots by correlating photoluminescence data with atomic force microscopy (AFM) images and positioning the MNPs in the vicinity of the QD using the AFM tip as was done by Ratchford \emph{et al.}~\cite{doi:10.1021/nl103906f}; in their study, the relatively small QDs 
had an estimated  dipole moment of around 5.3~Debye 
and a strong modification of the spontaneous emission rate was shown, along with a drastic reduction in blinking. A second possible method for probing the emission spectra of coupled QD-MNP systems could use an array of MNPs placed on a substrate and immersed in a solution of colloidal QDs. By illuminating with focussed (off-resonant) laser beams it is also possible to create efficient optical traps~\cite{Grigorenko-NATP-2-365W,chang:123004,Righini-Nat.Phys.-3-477} at which point the QD can be loaded into an excited state. Both of these proposals involve the use of substrates; of note our formalism enables the calculation of the far field spectrum in any inhomogeneous geometry as long as the Green function can be calculated; our initial results using the finite-difference
time-domain technique [see Fig.~\ref{fig:schematic}(b)] for this simplified geometry can easily include a substrate;  or  even more complicated geometries could be investigated such as MNPs coupled directly to photonic crystal cavities containing QDs~\cite{Barth-Nano.Lett.-10-891,Benson2011}.
In fact, recent experiments with QDs coupled with disordered metallic films on glass substrates are at a loss for the expected far field emission spectra~\cite{PhysRevB.84.245423}, further emphasizing the usefulness of our technique.

\section{Conclusions}
\label{sec:conclusions}
We have presented a 
Green-function quantum optics approach to study quantum optical interactions
between a dipole emitter and a single MNP. We began by examining the properties of the classical Green function above a MNP within and beyond the dipole approximation and showed the dramatic effects  of the higher order plasmon modes on the LDOS and photonic Lamb shifts. Going beyond the weak coupling approximation, we 
 then  examined the  particle spectrum and contrasted this with the far field (observable) light-emission spectrum of a QD coupled to the MNP. Using 
experimentally accessible parameters, our non-perturbative light spectra
 show clear signatures of the strong coupling regime;  the
emitted spectrum was found to contain a triplet or quartet of
resonances, highlighting the important role of light propagation to the detector.
Finally, we also examined the role of quenching on the far field spectra, and compared the quenching to the case of a metal half space. Our techniques are quite general and  can be extended to include an initial pump field, multiple MNPs and multiple QDs. 

\section*{Acknowledgements}
This work was supported by National Science and Research Council of Canada and The Danish Council for Independent Research (FTP 10-093651).

\begin{widetext}
\appendix*
\section{Spherical Green function}
 Given a sphere with permittivity $\varepsilon_m$ and radius $a$, embedded in a homogeneous medium of permittivity $\varepsilon_b$,  the scattered part of the Green function is given by
\begin{equation}
\label{eq:G_anal}
\begin{split}
 \GFT^{\rm scatt}\! \left(\mr,\mr'\right)\! =& \!\frac{-ik_b}{4\pi}\! \sum_{e,o} \!\sum_{\,n=1}^\infty\! \sum_{\,m=0}^n\! \left(2 - \delta_m^0 \right)\!\! \frac{2n+1}{n\left(n+1\right)} \frac{\left(n-m\right)!}{\left(n+m\right)!}
\!\!\left[R^H \mathbf{M}_{mn}^{eo}\!\left(k_b \mr\right) \!\mathbf{M}_{mn}^{eo}\!\left(k_b \mr' \right) +R_V \mathbf{N}_{mn}^{eo}\!\left(k_b \mr\right)\! \mathbf{N}_{mn}^{eo}\!\left(k_b \mr'\right)\right],
\end{split}
\end{equation}
where $R_H$/$R_V$ are the centrifugal reflection coefficients corresponding to transverse electric/magnetic waves (TE/TM), and $\mathbf{M}_{mn}^{eo}$/$\mathbf{N}_{mn}^{eo}$ are the vector functions corresponding to TE/TM waves and they have been separated into even and odd contributions. The values of $R_H$, $R_V$, are given by
\begin{align}
R^H = \frac{k_m \partial \tau_{m} \tau_{b} - k_b \partial \tau_{b} \tau_{m} }{k_m \partial \tau_{m} \kappa_{b} - k_b \partial \kappa_{b} \tau_{m} },
  \ \ \ \ R^V &^= \frac{k_m \tau_{m} \partial \tau_{b} - k_b \tau_{b} \partial \tau_{m} }{k_m  \tau_{m} \partial \kappa_{b} - k_b \kappa_{b} \partial \tau_{m} } ,
\end{align}
where
\begin{align}
 \tau_{i}  = j_n\left(k_i a\right), \quad \quad
 \kappa_{i}  = h^{(1)}_n\left(k_i a\right),
\end{align}
\begin{align}
 \partial \tau_{i}  =\frac{1}{k_i a} \frac{\partial (k_i a j_n\left(k_i a\right))}{ \partial 
 k_i a},
\ \ \
 \kappa_{i}  = \frac{1}{k_i a} \frac{\partial (k_i a h^{(1)}_n\left(k_i a\right))}{ \partial k_i a}.
\end{align}
Here $j_n$, $h^{(1)}_n$ are the spherical Bessel functions and spherical Hankel functions of the first kind respectively. The vector functions are defined as follows:
\begin{align}
\label{eq:Mvec_even}
\begin{split}
\mathbf{M}_{mn}^{e}\left(k \mr\right) =  - \frac{m}{\sin \theta } h^{(1)}\left(kr\right)P^m_n\left(\cos \theta\right) \sin m \phi \hat{\theta} 
 - h^{(1)}\left(kr\right) \frac{d P^m_n\left(\cos \theta\right)}{d \theta}  \cos m \phi \hat{\phi} ,
\end{split}
\end{align}
\begin{align}
\label{eq:Mvec_odd}
\begin{split}
\mathbf{M}_{mn}^{o}\left(k \mr\right) =  \frac{m}{\sin \theta } h^{(1)}\left(kr\right)P^m_n\left(\cos \theta\right) \cos m \phi \hat{\theta} 
& - h^{(1)}\left(kr\right) \frac{d P^m_n\left(\cos \theta\right)}{d \theta}  \sin m \phi \hat{\phi},
\end{split}
\end{align}
\begin{align}
\label{eq:Nvec_even}
\mathbf{N}_{mn}^{e}\left(k \mr\right) &=  \frac{n\left(n+1\right)}{kr}  h^{(1)}\left(kr\right)P^m_n\left(\cos \theta\right) \cos m \phi \hat{r} \nonumber \\
&+ \frac{1}{kr} \frac{d (r  h^{(1)}\left(kr\right))}{d r} \left[\frac{d P^m_n\left(\cos \theta\right)}{d \theta} \cos m \phi \hat{\theta} 
 - \frac{m}{\sin \theta } P^m_n\left(\cos \theta\right)  \sin m \phi \hat{\phi} \right]
\end{align}
\begin{align}
\label{eq:Nvec_odd}
\mathbf{N}_{mn}^{o}\left(k \mr\right) &=  \frac{n\left(n+1\right)}{kr}  h^{(1)}\left(kr\right)P^m_n\left(\cos \theta\right) \sin m \phi \hat{r} \nonumber \\
&+ \frac{1}{kr} \frac{d (r  h^{(1)}\left(kr\right))}{d r} \left[\frac{d P^m_n\left(\cos \theta\right)}{d \theta} \sin m \phi \hat{\theta} + \frac{m}{\sin \theta } P^m_n\left(\cos \theta\right)  \cos m \phi \hat{\phi} \right].
\end{align}
Note that for our numerical calculation in this paper,  a few simplifying assumptions can be made; we only consider the calculation to be along the $z$ direction, $x=x'=y=y'=0$, and we additionally assume that we are only calculating the LDOS ($z=z'$). This means that calculating in the $\hat{\theta}\hat{\theta}$ direction is equivalent to the $\hat{\phi}\hat{\phi}$ direction. This allows us to simplify Eqs.~(\ref{eq:Mvec_even})-(\ref{eq:Nvec_odd}) to,
\begin{align}
\label{eq:Mvec_even_simp}
\begin{split}
\mathbf{M}_{mn}^{e}\left(k \mr\right) =&  h^{(1)}\left(kr\right) \frac{d P^m_n\left(0 \right)}{d \theta} \hat{\phi},
\end{split}
\end{align}
\begin{align}
\label{eq:Mvec_odd_simp}
\begin{split}
\mathbf{M}_{mn}^{o}\left(k \mr\right) =&  0,
\end{split}
\end{align}
\begin{align}
\label{eq:Nvec_even_simp}
\begin{split}
\mathbf{N}_{mn}^{e}\left(k \mr\right) =&  \frac{n\left(n+1\right)}{kr}  h^{(1)}\left(kr\right)P^m_n\left(0 \right) \hat{r},
\end{split}
\end{align}
\begin{align}
\label{eq:Nvec_odd_simp}
\begin{split}
\mathbf{N}_{mn}^{o}\left(k \mr\right) =& \frac{m}{kr} \frac{d (r  h^{(1)}\left(kr\right))}{d r} P^m_n\left(0 \right) \hat{\phi}.
\end{split}
\end{align}

\end{widetext}


\begin{thebibliography}{45}%
\makeatletter
\providecommand \@ifxundefined [1]{%
 \@ifx{#1\undefined}
}%
\providecommand \@ifnum [1]{%
 \ifnum #1\expandafter \@firstoftwo
 \else \expandafter \@secondoftwo
 \fi
}%
\providecommand \@ifx [1]{%
 \ifx #1\expandafter \@firstoftwo
 \else \expandafter \@secondoftwo
 \fi
}%
\providecommand \natexlab [1]{#1}%
\providecommand \enquote  [1]{``#1''}%
\providecommand \bibnamefont  [1]{#1}%
\providecommand \bibfnamefont [1]{#1}%
\providecommand \citenamefont [1]{#1}%
\providecommand \href@noop [0]{\@secondoftwo}%
\providecommand \href [0]{\begingroup \@sanitize@url \@href}%
\providecommand \@href[1]{\@@startlink{#1}\@@href}%
\providecommand \@@href[1]{\endgroup#1\@@endlink}%
\providecommand \@sanitize@url [0]{\catcode `\\12\catcode `\$12\catcode
  `\&12\catcode `\#12\catcode `\^12\catcode `\_12\catcode `\%12\relax}%
\providecommand \@@startlink[1]{}%
\providecommand \@@endlink[0]{}%
\providecommand \url  [0]{\begingroup\@sanitize@url \@url }%
\providecommand \@url [1]{\endgroup\@href {#1}{\urlprefix }}%
\providecommand \urlprefix  [0]{URL }%
\providecommand \Eprint [0]{\href }%
\providecommand \doibase [0]{http://dx.doi.org/}%
\providecommand \selectlanguage [0]{\@gobble}%
\providecommand \bibinfo  [0]{\@secondoftwo}%
\providecommand \bibfield  [0]{\@secondoftwo}%
\providecommand \translation [1]{[#1]}%
\providecommand \BibitemOpen [0]{}%
\providecommand \bibitemStop [0]{}%
\providecommand \bibitemNoStop [0]{.\EOS\space}%
\providecommand \EOS [0]{\spacefactor3000\relax}%
\providecommand \BibitemShut  [1]{\csname bibitem#1\endcsname}%
\let\auto@bib@innerbib\@empty
\bibitem [{\citenamefont {Vahala}(2003)}]{Vahala2003}%
  \BibitemOpen
  \bibfield  {author} {\bibinfo {author} {\bibfnamefont {K.~J.}\ \bibnamefont
  {Vahala}},\ }\href {\doibase 10.1038/nature01939} {\bibfield  {journal}
  {\bibinfo  {journal} {Nature}\ }\textbf {\bibinfo {volume} {424}},\ \bibinfo
  {pages} {839} (\bibinfo {year} {2003})}\BibitemShut {NoStop}%
\bibitem [{\citenamefont {Purcell}(1946)}]{Purcell-PR-69-681}%
  \BibitemOpen
  \bibfield  {author} {\bibinfo {author} {\bibfnamefont {E.~M.}\ \bibnamefont
  {Purcell}},\ }\href@noop {} {\bibfield  {journal} {\bibinfo  {journal} {Phys.
  Rev.}\ }\textbf {\bibinfo {volume} {69}},\ \bibinfo {pages} {681} (\bibinfo
  {year} {1946})}\BibitemShut {NoStop}%
\bibitem [{\citenamefont {Akahane}\ \emph {et~al.}(2003)\citenamefont
  {Akahane}, \citenamefont {Asano}, \citenamefont {Song},\ and\ \citenamefont
  {Noda}}]{Akahane-NAT-425-944}%
  \BibitemOpen
  \bibfield  {author} {\bibinfo {author} {\bibfnamefont {Y.}~\bibnamefont
  {Akahane}}, \bibinfo {author} {\bibfnamefont {T.}~\bibnamefont {Asano}},
  \bibinfo {author} {\bibfnamefont {B.-S.}\ \bibnamefont {Song}}, \ and\
  \bibinfo {author} {\bibfnamefont {S.}~\bibnamefont {Noda}},\ }\href {\doibase
  10.1038/nature02063} {\bibfield  {journal} {\bibinfo  {journal} {Nature}\
  }\textbf {\bibinfo {volume} {425}},\ \bibinfo {pages} {944} (\bibinfo {year}
  {2003})}\BibitemShut {NoStop}%
\bibitem [{\citenamefont {Reithmaier}\ \emph {et~al.}(2004)\citenamefont
  {Reithmaier}, \citenamefont {Sek}, \citenamefont {L\"offler}, \citenamefont
  {Hofmann}, \citenamefont {Kuhn}, \citenamefont {Reitzenstein}, \citenamefont
  {Keldysh}, \citenamefont {Kulakovskii}, \citenamefont {Reinecke},\ and\
  \citenamefont {Forchel}}]{qdQED}%
  \BibitemOpen
  \bibfield  {author} {\bibinfo {author} {\bibfnamefont {J.~P.}\ \bibnamefont
  {Reithmaier}}, \bibinfo {author} {\bibfnamefont {G.}~\bibnamefont {Sek}},
  \bibinfo {author} {\bibfnamefont {A.}~\bibnamefont {L\"offler}}, \bibinfo
  {author} {\bibfnamefont {C.}~\bibnamefont {Hofmann}}, \bibinfo {author}
  {\bibfnamefont {S.}~\bibnamefont {Kuhn}}, \bibinfo {author} {\bibfnamefont
  {S.}~\bibnamefont {Reitzenstein}}, \bibinfo {author} {\bibfnamefont {L.~V.}\
  \bibnamefont {Keldysh}}, \bibinfo {author} {\bibfnamefont {V.~D.}\
  \bibnamefont {Kulakovskii}}, \bibinfo {author} {\bibfnamefont {T.~L.}\
  \bibnamefont {Reinecke}}, \ and\ \bibinfo {author} {\bibfnamefont
  {A.}~\bibnamefont {Forchel}},\ }\href {\doibase 10.1038/nature02969}
  {\bibfield  {journal} {\bibinfo  {journal} {Nature}\ }\textbf {\bibinfo
  {volume} {432}},\ \bibinfo {pages} {197} (\bibinfo {year}
  {2004})}\BibitemShut {NoStop}%
\bibitem [{\citenamefont {Maier}(2007)}]{Maier-Plamonics}%
  \BibitemOpen
  \bibfield  {author} {\bibinfo {author} {\bibfnamefont {S.~A.}\ \bibnamefont
  {Maier}},\ }\href@noop {} {\emph {\bibinfo {title} {Plasmonics: Fundamentals
  and Applications}}}\ (\bibinfo  {publisher} {Springer US},\ \bibinfo {year}
  {2007})\BibitemShut {NoStop}%
\bibitem [{\citenamefont {Ratchford}\ \emph {et~al.}(2011)\citenamefont
  {Ratchford}, \citenamefont {Shafiei}, \citenamefont {Kim}, \citenamefont
  {Gray},\ and\ \citenamefont {Li}}]{doi:10.1021/nl103906f}%
  \BibitemOpen
  \bibfield  {author} {\bibinfo {author} {\bibfnamefont {D.}~\bibnamefont
  {Ratchford}}, \bibinfo {author} {\bibfnamefont {F.}~\bibnamefont {Shafiei}},
  \bibinfo {author} {\bibfnamefont {S.}~\bibnamefont {Kim}}, \bibinfo {author}
  {\bibfnamefont {S.~K.}\ \bibnamefont {Gray}}, \ and\ \bibinfo {author}
  {\bibfnamefont {X.}~\bibnamefont {Li}},\ }\href {\doibase 10.1021/nl103906f}
  {\bibfield  {journal} {\bibinfo  {journal} {Nano Letters}\ }\textbf {\bibinfo
  {volume} {11}},\ \bibinfo {pages} {1049} (\bibinfo {year}
  {2011})}\BibitemShut {NoStop}%
\bibitem [{\citenamefont {Savasta}\ \emph {et~al.}(2010)\citenamefont
  {Savasta}, \citenamefont {Saija}, \citenamefont {Ridolfo}, \citenamefont
  {Di~Stefano}, \citenamefont {Denti},\ and\ \citenamefont
  {Borghese}}]{doi:10.1021/nn100585h}%
  \BibitemOpen
  \bibfield  {author} {\bibinfo {author} {\bibfnamefont {S.}~\bibnamefont
  {Savasta}}, \bibinfo {author} {\bibfnamefont {R.}~\bibnamefont {Saija}},
  \bibinfo {author} {\bibfnamefont {A.}~\bibnamefont {Ridolfo}}, \bibinfo
  {author} {\bibfnamefont {O.}~\bibnamefont {Di~Stefano}}, \bibinfo {author}
  {\bibfnamefont {P.}~\bibnamefont {Denti}}, \ and\ \bibinfo {author}
  {\bibfnamefont {F.}~\bibnamefont {Borghese}},\ }\href {\doibase
  10.1021/nn100585h} {\bibfield  {journal} {\bibinfo  {journal} {ACS Nano}\
  }\textbf {\bibinfo {volume} {4}},\ \bibinfo {pages} {6369} (\bibinfo {year}
  {2010})}\BibitemShut {NoStop}%
\bibitem [{\citenamefont {Duan}\ and\ \citenamefont
  {Kimble}(2004)}]{PhysRevLett.92.127902}%
  \BibitemOpen
  \bibfield  {author} {\bibinfo {author} {\bibfnamefont {L.-M.}\ \bibnamefont
  {Duan}}\ and\ \bibinfo {author} {\bibfnamefont {H.~J.}\ \bibnamefont
  {Kimble}},\ }\href {\doibase 10.1103/PhysRevLett.92.127902} {\bibfield
  {journal} {\bibinfo  {journal} {Phys. Rev. Lett.}\ }\textbf {\bibinfo
  {volume} {92}},\ \bibinfo {pages} {127902} (\bibinfo {year}
  {2004})}\BibitemShut {NoStop}%
\bibitem [{\citenamefont {Monroe}(2002)}]{Monroe2002}%
  \BibitemOpen
  \bibfield  {author} {\bibinfo {author} {\bibfnamefont {C.}~\bibnamefont
  {Monroe}},\ }\href {\doibase 10.1038/416238a} {\bibfield  {journal} {\bibinfo
   {journal} {Nature}\ }\textbf {\bibinfo {volume} {416}},\ \bibinfo {pages}
  {238} (\bibinfo {year} {2002})}\BibitemShut {NoStop}%
\bibitem [{\citenamefont {Bergman}\ and\ \citenamefont
  {Stockman}(2003)}]{PhysRevLett.90.027402}%
  \BibitemOpen
  \bibfield  {author} {\bibinfo {author} {\bibfnamefont {D.~J.}\ \bibnamefont
  {Bergman}}\ and\ \bibinfo {author} {\bibfnamefont {M.~I.}\ \bibnamefont
  {Stockman}},\ }\href {\doibase 10.1103/PhysRevLett.90.027402} {\bibfield
  {journal} {\bibinfo  {journal} {Phys. Rev. Lett.}\ }\textbf {\bibinfo
  {volume} {90}},\ \bibinfo {pages} {027402} (\bibinfo {year}
  {2003})}\BibitemShut {NoStop}%
\bibitem [{\citenamefont {Noginov}\ \emph {et~al.}(2009)\citenamefont
  {Noginov}, \citenamefont {Zhu}, \citenamefont {Belgrave}, \citenamefont
  {Bakker}, \citenamefont {Shalaev}, \citenamefont {Narimanov}, \citenamefont
  {Stout}, \citenamefont {Herz}, \citenamefont {Suteewong},\ and\ \citenamefont
  {Wiesner}}]{Noginov-Nature-460-1110}%
  \BibitemOpen
  \bibfield  {author} {\bibinfo {author} {\bibfnamefont {M.~A.}\ \bibnamefont
  {Noginov}}, \bibinfo {author} {\bibfnamefont {G.}~\bibnamefont {Zhu}},
  \bibinfo {author} {\bibfnamefont {A.~M.}\ \bibnamefont {Belgrave}}, \bibinfo
  {author} {\bibfnamefont {R.}~\bibnamefont {Bakker}}, \bibinfo {author}
  {\bibfnamefont {V.~M.}\ \bibnamefont {Shalaev}}, \bibinfo {author}
  {\bibfnamefont {E.~E.}\ \bibnamefont {Narimanov}}, \bibinfo {author}
  {\bibfnamefont {S.}~\bibnamefont {Stout}}, \bibinfo {author} {\bibfnamefont
  {E.}~\bibnamefont {Herz}}, \bibinfo {author} {\bibfnamefont {T.}~\bibnamefont
  {Suteewong}}, \ and\ \bibinfo {author} {\bibfnamefont {U.}~\bibnamefont
  {Wiesner}},\ }\href {\doibase 10.1038/nature08318} {\bibfield  {journal}
  {\bibinfo  {journal} {Nature}\ }\textbf {\bibinfo {volume} {460}},\ \bibinfo
  {pages} {1110} (\bibinfo {year} {2009})}\BibitemShut {NoStop}%
\bibitem [{\citenamefont {Zheludev}\ \emph {et~al.}(2008)\citenamefont
  {Zheludev}, \citenamefont {Prosvirnin}, \citenamefont {Papasimakis},\ and\
  \citenamefont {Fedotov}}]{Zheludev-NaturePhotonics-2-351}%
  \BibitemOpen
  \bibfield  {author} {\bibinfo {author} {\bibfnamefont {N.~I.}\ \bibnamefont
  {Zheludev}}, \bibinfo {author} {\bibfnamefont {S.~L.}\ \bibnamefont
  {Prosvirnin}}, \bibinfo {author} {\bibfnamefont {N.}~\bibnamefont
  {Papasimakis}}, \ and\ \bibinfo {author} {\bibfnamefont {V.~A.}\ \bibnamefont
  {Fedotov}},\ }\href {\doibase 10.1038/nphoton.2008.82} {\bibfield  {journal}
  {\bibinfo  {journal} {Nature Photonics}\ }\textbf {\bibinfo {volume} {2}},\
  \bibinfo {pages} {351} (\bibinfo {year} {2008})}\BibitemShut {NoStop}%
\bibitem [{\citenamefont {Waks}\ and\ \citenamefont
  {Sridharan}(2010)}]{PhysRevA.82.043845}%
  \BibitemOpen
  \bibfield  {author} {\bibinfo {author} {\bibfnamefont {E.}~\bibnamefont
  {Waks}}\ and\ \bibinfo {author} {\bibfnamefont {D.}~\bibnamefont
  {Sridharan}},\ }\href {\doibase 10.1103/PhysRevA.82.043845} {\bibfield
  {journal} {\bibinfo  {journal} {Phys. Rev. A}\ }\textbf {\bibinfo {volume}
  {82}},\ \bibinfo {pages} {043845} (\bibinfo {year} {2010})}\BibitemShut
  {NoStop}%
\bibitem [{\citenamefont {Carminati}\ \emph {et~al.}(2006)\citenamefont
  {Carminati}, \citenamefont {Greffet}, \citenamefont {Henkel},\ and\
  \citenamefont {Vigoureux}}]{Carminati2006368}%
  \BibitemOpen
  \bibfield  {author} {\bibinfo {author} {\bibfnamefont {R.}~\bibnamefont
  {Carminati}}, \bibinfo {author} {\bibfnamefont {J.-J.}\ \bibnamefont
  {Greffet}}, \bibinfo {author} {\bibfnamefont {C.}~\bibnamefont {Henkel}}, \
  and\ \bibinfo {author} {\bibfnamefont {J.}~\bibnamefont {Vigoureux}},\ }\href
  {\doibase DOI: 10.1016/j.optcom.2005.12.009} {\bibfield  {journal} {\bibinfo
  {journal} {Optics Communications}\ }\textbf {\bibinfo {volume} {261}},\
  \bibinfo {pages} {368 } (\bibinfo {year} {2006})}\BibitemShut {NoStop}%
\bibitem [{\citenamefont {Castani\'{e}}\ \emph {et~al.}(2010)\citenamefont
  {Castani\'{e}}, \citenamefont {Boffety},\ and\ \citenamefont
  {Carminati}}]{Castanie:10}%
  \BibitemOpen
  \bibfield  {author} {\bibinfo {author} {\bibfnamefont {E.}~\bibnamefont
  {Castani\'{e}}}, \bibinfo {author} {\bibfnamefont {M.}~\bibnamefont
  {Boffety}}, \ and\ \bibinfo {author} {\bibfnamefont {R.}~\bibnamefont
  {Carminati}},\ }\href {\doibase 10.1364/OL.35.000291} {\bibfield  {journal}
  {\bibinfo  {journal} {Opt. Lett.}\ }\textbf {\bibinfo {volume} {35}},\
  \bibinfo {pages} {291} (\bibinfo {year} {2010})}\BibitemShut {NoStop}%
\bibitem [{\citenamefont {Anger}\ \emph {et~al.}(2006)\citenamefont {Anger},
  \citenamefont {Bharadwaj},\ and\ \citenamefont
  {Novotny}}]{PhysRevLett.96.113002}%
  \BibitemOpen
  \bibfield  {author} {\bibinfo {author} {\bibfnamefont {P.}~\bibnamefont
  {Anger}}, \bibinfo {author} {\bibfnamefont {P.}~\bibnamefont {Bharadwaj}}, \
  and\ \bibinfo {author} {\bibfnamefont {L.}~\bibnamefont {Novotny}},\ }\href
  {\doibase 10.1103/PhysRevLett.96.113002} {\bibfield  {journal} {\bibinfo
  {journal} {Phys. Rev. Lett.}\ }\textbf {\bibinfo {volume} {96}},\ \bibinfo
  {pages} {113002} (\bibinfo {year} {2006})}\BibitemShut {NoStop}%
\bibitem [{\citenamefont {Ruppin}(1982)}]{10.1063/1.443196}%
  \BibitemOpen
  \bibfield  {author} {\bibinfo {author} {\bibfnamefont {R.}~\bibnamefont
  {Ruppin}},\ }\href {\doibase DOI:10.1063/1.443196} {\bibfield  {journal}
  {\bibinfo  {journal} {J. Chem. Phys.}\ }\textbf {\bibinfo {volume} {76}},\
  \bibinfo {pages} {1681} (\bibinfo {year} {1982})}\BibitemShut {NoStop}%
\bibitem [{\citenamefont {Tr\"ugler}\ and\ \citenamefont
  {Hohenester}(2008)}]{PhysRevB.77.115403}%
  \BibitemOpen
  \bibfield  {author} {\bibinfo {author} {\bibfnamefont {A.}~\bibnamefont
  {Tr\"ugler}}\ and\ \bibinfo {author} {\bibfnamefont {U.}~\bibnamefont
  {Hohenester}},\ }\href {\doibase 10.1103/PhysRevB.77.115403} {\bibfield
  {journal} {\bibinfo  {journal} {Phys. Rev. B}\ }\textbf {\bibinfo {volume}
  {77}},\ \bibinfo {pages} {115403} (\bibinfo {year} {2008})}\BibitemShut
  {NoStop}%
\bibitem [{\citenamefont {Hohenester}\ and\ \citenamefont
  {Trugler}(2008)}]{4683624}%
  \BibitemOpen
  \bibfield  {author} {\bibinfo {author} {\bibfnamefont {U.}~\bibnamefont
  {Hohenester}}\ and\ \bibinfo {author} {\bibfnamefont {A.}~\bibnamefont
  {Trugler}},\ }\href {\doibase 10.1109/JSTQE.2008.2007918} {\bibfield
  {journal} {\bibinfo  {journal} {Selected Topics in Quantum Electronics, IEEE
  Journal of}\ }\textbf {\bibinfo {volume} {14}},\ \bibinfo {pages} {1430 }
  (\bibinfo {year} {2008})}\BibitemShut {NoStop}%
\bibitem [{\citenamefont {Carmichael}(1999)}]{Carmichael1999}%
  \BibitemOpen
  \bibfield  {author} {\bibinfo {author} {\bibfnamefont {H.~J.}\ \bibnamefont
  {Carmichael}},\ }\href@noop {} {\emph {\bibinfo {title} {Statistical Methods
  in Quantum Optics 1}}}\ (\bibinfo  {publisher} {Springer-Verlag},\ \bibinfo
  {year} {1999})\BibitemShut {NoStop}%
\bibitem [{\citenamefont {Tai}(1971)}]{Tai-GF}%
  \BibitemOpen
  \bibfield  {author} {\bibinfo {author} {\bibfnamefont {C.-T.}\ \bibnamefont
  {Tai}},\ }\href@noop {} {\emph {\bibinfo {title} {Dyadic Green's Functions in
  Electromagnetic Theory}}},\ edited by\ \bibinfo {editor} {\bibfnamefont
  {D.~K.}\ \bibnamefont {Cheng}}\ (\bibinfo  {publisher} {Intext Educational
  Publishers},\ \bibinfo {year} {1971})\BibitemShut {NoStop}%
\bibitem [{\citenamefont {Li}\ \emph {et~al.}(1994)\citenamefont {Li},
  \citenamefont {Kooi}, \citenamefont {Leong},\ and\ \citenamefont
  {Yeo}}]{L.-W.Li1994}%
  \BibitemOpen
  \bibfield  {author} {\bibinfo {author} {\bibfnamefont {L.-W.}\ \bibnamefont
  {Li}}, \bibinfo {author} {\bibfnamefont {P.-S.}\ \bibnamefont {Kooi}},
  \bibinfo {author} {\bibfnamefont {M.-S.}\ \bibnamefont {Leong}}, \ and\
  \bibinfo {author} {\bibfnamefont {T.-S.}\ \bibnamefont {Yeo}},\ }\href
  {\doibase 10.1109/22.339756} {\bibfield  {journal} {\bibinfo  {journal} {IEEE
  Trans. Micro. Theory and Tech.}\ }\textbf {\bibinfo {volume} {42}},\ \bibinfo
  {pages} {2302} (\bibinfo {year} {1994})}\BibitemShut {NoStop}%
\bibitem [{\citenamefont {Draine}(1988)}]{Draine-AstrophysJ.-333-848}%
  \BibitemOpen
  \bibfield  {author} {\bibinfo {author} {\bibfnamefont {B.}~\bibnamefont
  {Draine}},\ }\href@noop {} {\bibfield  {journal} {\bibinfo  {journal}
  {Astrophys J.}\ }\textbf {\bibinfo {volume} {333}},\ \bibinfo {pages} {848}
  (\bibinfo {year} {1988})}\BibitemShut {NoStop}%
\bibitem [{\citenamefont {Greffet}\ \emph {et~al.}(2010)\citenamefont
  {Greffet}, \citenamefont {Laroche},\ and\ \citenamefont
  {Marquier}}]{PhysRevLett.105.117701}%
  \BibitemOpen
  \bibfield  {author} {\bibinfo {author} {\bibfnamefont {J.-J.}\ \bibnamefont
  {Greffet}}, \bibinfo {author} {\bibfnamefont {M.}~\bibnamefont {Laroche}}, \
  and\ \bibinfo {author} {\bibfnamefont {F.}~\bibnamefont {Marquier}},\ }\href
  {\doibase 10.1103/PhysRevLett.105.117701} {\bibfield  {journal} {\bibinfo
  {journal} {Phys. Rev. Lett.}\ }\textbf {\bibinfo {volume} {105}},\ \bibinfo
  {pages} {117701} (\bibinfo {year} {2010})}\BibitemShut {NoStop}%
\bibitem [{\citenamefont {Novotny}\ and\ \citenamefont
  {Hecht}(2006)}]{Novotny-Nano-Optics}%
  \BibitemOpen
  \bibfield  {author} {\bibinfo {author} {\bibfnamefont {L.}~\bibnamefont
  {Novotny}}\ and\ \bibinfo {author} {\bibfnamefont {B.}~\bibnamefont
  {Hecht}},\ }\href@noop {} {\emph {\bibinfo {title} {Principles of
  Nano-Optics}}}\ (\bibinfo  {publisher} {Cambridge},\ \bibinfo {year}
  {2006})\BibitemShut {NoStop}%
\bibitem [{\citenamefont {Yao}\ \emph {et~al.}(2009)\citenamefont {Yao},
  \citenamefont {Van~Vlack}, \citenamefont {Reza}, \citenamefont {Patterson},
  \citenamefont {Dignam},\ and\ \citenamefont
  {Hughes}}]{Yao-PhysicalReviewB-80-195106}%
  \BibitemOpen
  \bibfield  {author} {\bibinfo {author} {\bibfnamefont {P.}~\bibnamefont
  {Yao}}, \bibinfo {author} {\bibfnamefont {C.}~\bibnamefont {Van~Vlack}},
  \bibinfo {author} {\bibfnamefont {A.}~\bibnamefont {Reza}}, \bibinfo {author}
  {\bibfnamefont {M.}~\bibnamefont {Patterson}}, \bibinfo {author}
  {\bibfnamefont {M.~M.}\ \bibnamefont {Dignam}}, \ and\ \bibinfo {author}
  {\bibfnamefont {S.}~\bibnamefont {Hughes}},\ }\href {\doibase
  10.1103/PhysRevB.80.195106} {\bibfield  {journal} {\bibinfo  {journal} {Phys.
  Rev. B}\ }\textbf {\bibinfo {volume} {80}},\ \bibinfo {eid} {195106}
  (\bibinfo {year} {2009})}\BibitemShut {NoStop}%
\bibitem [{\citenamefont {Dung}\ \emph {et~al.}(1998)\citenamefont {Dung},
  \citenamefont {Kn\"oll},\ and\ \citenamefont {Welsch}}]{Dung-PRA-57-3931}%
  \BibitemOpen
  \bibfield  {author} {\bibinfo {author} {\bibfnamefont {H.~T.}\ \bibnamefont
  {Dung}}, \bibinfo {author} {\bibfnamefont {L.}~\bibnamefont {Kn\"oll}}, \
  and\ \bibinfo {author} {\bibfnamefont {D.-G.}\ \bibnamefont {Welsch}},\
  }\href {\doibase 10.1103/PhysRevA.57.3931} {\bibfield  {journal} {\bibinfo
  {journal} {Phys. Rev. A}\ }\textbf {\bibinfo {volume} {57}},\ \bibinfo
  {pages} {3931} (\bibinfo {year} {1998})}\BibitemShut {NoStop}%
\bibitem [{\citenamefont {Dung}\ \emph {et~al.}(2003)\citenamefont {Dung},
  \citenamefont {Buhmann}, \citenamefont {Kn\"oll}, \citenamefont {Welsch},
  \citenamefont {Scheel},\ and\ \citenamefont {K\"astel}}]{Dung-PRA-68-043816}%
  \BibitemOpen
  \bibfield  {author} {\bibinfo {author} {\bibfnamefont {H.}~\bibnamefont
  {Dung}}, \bibinfo {author} {\bibfnamefont {S.}~\bibnamefont {Buhmann}},
  \bibinfo {author} {\bibfnamefont {L.}~\bibnamefont {Kn\"oll}}, \bibinfo
  {author} {\bibfnamefont {D.}~\bibnamefont {Welsch}}, \bibinfo {author}
  {\bibfnamefont {S.}~\bibnamefont {Scheel}}, \ and\ \bibinfo {author}
  {\bibfnamefont {J.}~\bibnamefont {K\"astel}},\ }\href {\doibase
  10.1103/PhysRevA.68.043816} {\bibfield  {journal} {\bibinfo  {journal} {Phys.
  Rev. A}\ }\textbf {\bibinfo {volume} {68}},\ \bibinfo {pages} {043816}
  (\bibinfo {year} {2003})}\BibitemShut {NoStop}%
\bibitem [{\citenamefont {Vogel}\ \emph {et~al.}(2006)\citenamefont {Vogel}, ,\
  and\ \citenamefont {Welsch}}]{Vogel-Quantum-Optics}%
  \BibitemOpen
  \bibfield  {author} {\bibinfo {author} {\bibfnamefont {W.}~\bibnamefont
  {Vogel}}, , \ and\ \bibinfo {author} {\bibfnamefont {G.}~\bibnamefont
  {Welsch}},\ }\href@noop {} {\emph {\bibinfo {title} {Quantum Optics}}}\
  (\bibinfo  {publisher} {Wiley-VCH},\ \bibinfo {year} {2006})\BibitemShut
  {NoStop}%
\bibitem [{\citenamefont {Kristensen}\ \emph {et~al.}(2011)\citenamefont
  {Kristensen}, \citenamefont {M\o{}rk}, \citenamefont {Lodahl},\ and\
  \citenamefont {Hughes}}]{PhysRevB.83.075305}%
  \BibitemOpen
  \bibfield  {author} {\bibinfo {author} {\bibfnamefont {P.~T.}\ \bibnamefont
  {Kristensen}}, \bibinfo {author} {\bibfnamefont {J.}~\bibnamefont {M\o{}rk}},
  \bibinfo {author} {\bibfnamefont {P.}~\bibnamefont {Lodahl}}, \ and\ \bibinfo
  {author} {\bibfnamefont {S.}~\bibnamefont {Hughes}},\ }\href {\doibase
  10.1103/PhysRevB.83.075305} {\bibfield  {journal} {\bibinfo  {journal} {Phys.
  Rev. B}\ }\textbf {\bibinfo {volume} {83}},\ \bibinfo {pages} {075305}
  (\bibinfo {year} {2011})}\BibitemShut {NoStop}%
\bibitem [{\citenamefont {Hughes}\ \emph {et~al.}(2011)\citenamefont {Hughes},
  \citenamefont {Yao}, \citenamefont {Milde}, \citenamefont {Knorr},
  \citenamefont {Dalacu}, \citenamefont {Mnaymneh}, \citenamefont {Sazonova},
  \citenamefont {Poole}, \citenamefont {Aers}, \citenamefont {Lapointe},
  \citenamefont {Cheriton},\ and\ \citenamefont
  {Williams}}]{PhysRevB.83.165313}%
  \BibitemOpen
  \bibfield  {author} {\bibinfo {author} {\bibfnamefont {S.}~\bibnamefont
  {Hughes}}, \bibinfo {author} {\bibfnamefont {P.}~\bibnamefont {Yao}},
  \bibinfo {author} {\bibfnamefont {F.}~\bibnamefont {Milde}}, \bibinfo
  {author} {\bibfnamefont {A.}~\bibnamefont {Knorr}}, \bibinfo {author}
  {\bibfnamefont {D.}~\bibnamefont {Dalacu}}, \bibinfo {author} {\bibfnamefont
  {K.}~\bibnamefont {Mnaymneh}}, \bibinfo {author} {\bibfnamefont
  {V.}~\bibnamefont {Sazonova}}, \bibinfo {author} {\bibfnamefont {P.~J.}\
  \bibnamefont {Poole}}, \bibinfo {author} {\bibfnamefont {G.~C.}\ \bibnamefont
  {Aers}}, \bibinfo {author} {\bibfnamefont {J.}~\bibnamefont {Lapointe}},
  \bibinfo {author} {\bibfnamefont {R.}~\bibnamefont {Cheriton}}, \ and\
  \bibinfo {author} {\bibfnamefont {R.~L.}\ \bibnamefont {Williams}},\ }\href
  {\doibase 10.1103/PhysRevB.83.165313} {\bibfield  {journal} {\bibinfo
  {journal} {Phys. Rev. B}\ }\textbf {\bibinfo {volume} {83}},\ \bibinfo
  {pages} {165313} (\bibinfo {year} {2011})}\BibitemShut {NoStop}%
\bibitem [{\citenamefont {Qiao}\ \emph {et~al.}(2010)\citenamefont {Qiao},
  \citenamefont {Abel}, \citenamefont {van Veggel},\ and\ \citenamefont
  {Young}}]{PhysRevB.82.165435}%
  \BibitemOpen
  \bibfield  {author} {\bibinfo {author} {\bibfnamefont {H.}~\bibnamefont
  {Qiao}}, \bibinfo {author} {\bibfnamefont {K.~A.}\ \bibnamefont {Abel}},
  \bibinfo {author} {\bibfnamefont {F.~C. J.~M.}\ \bibnamefont {van Veggel}}, \
  and\ \bibinfo {author} {\bibfnamefont {J.~F.}\ \bibnamefont {Young}},\ }\href
  {\doibase 10.1103/PhysRevB.82.165435} {\bibfield  {journal} {\bibinfo
  {journal} {Phys. Rev. B}\ }\textbf {\bibinfo {volume} {82}},\ \bibinfo
  {pages} {165435} (\bibinfo {year} {2010})}\BibitemShut {NoStop}%
\bibitem [{\citenamefont {Wilson-Rae}\ and\ \citenamefont
  {Imamo\ifmmode~\breve{g}\else \u{g}\fi{}lu}(2002)}]{PhysRevB.65.235311}%
  \BibitemOpen
  \bibfield  {author} {\bibinfo {author} {\bibfnamefont {I.}~\bibnamefont
  {Wilson-Rae}}\ and\ \bibinfo {author} {\bibfnamefont {A.}~\bibnamefont
  {Imamo\ifmmode~\breve{g}\else \u{g}\fi{}lu}},\ }\href {\doibase
  10.1103/PhysRevB.65.235311} {\bibfield  {journal} {\bibinfo  {journal} {Phys.
  Rev. B}\ }\textbf {\bibinfo {volume} {65}},\ \bibinfo {pages} {235311}
  (\bibinfo {year} {2002})}\BibitemShut {NoStop}%
\bibitem [{\citenamefont {Roy}\ and\ \citenamefont
  {Hughes}(2011)}]{PhysRevLett.106.247403}%
  \BibitemOpen
  \bibfield  {author} {\bibinfo {author} {\bibfnamefont {C.}~\bibnamefont
  {Roy}}\ and\ \bibinfo {author} {\bibfnamefont {S.}~\bibnamefont {Hughes}},\
  }\href {\doibase 10.1103/PhysRevLett.106.247403} {\bibfield  {journal}
  {\bibinfo  {journal} {Phys. Rev. Lett.}\ }\textbf {\bibinfo {volume} {106}},\
  \bibinfo {pages} {247403} (\bibinfo {year} {2011})}\BibitemShut {NoStop}%
\bibitem [{lum()}]{lumerical}%
  \BibitemOpen
  \href@noop {} {}\bibinfo {howpublished} {We use Lumerical's FDTD Solutions: www.lumerical.com}\BibitemShut
  {NoStop}%
\bibitem [{\citenamefont {Sun}\ and\ \citenamefont
  {Khurgin}(2010)}]{10.1063/1.3532101}%
  \BibitemOpen
  \bibfield  {author} {\bibinfo {author} {\bibfnamefont {G.}~\bibnamefont
  {Sun}}\ and\ \bibinfo {author} {\bibfnamefont {J.~B.}\ \bibnamefont
  {Khurgin}},\ }\href {\doibase 10.1063/1.3532101} {\bibfield  {journal}
  {\bibinfo  {journal} {Appl. Phys. Lett.}\ }\textbf {\bibinfo {volume} {97}},\
  \bibinfo {pages} {263110} (\bibinfo {year} {2010})}\BibitemShut {NoStop}%
\bibitem [{\citenamefont {Wang}\ \emph {et~al.}(2004)\citenamefont {Wang},
  \citenamefont {Kivshar},\ and\ \citenamefont {Gu}}]{Wang-PRL-93-073901}%
  \BibitemOpen
  \bibfield  {author} {\bibinfo {author} {\bibfnamefont {X.-H.}\ \bibnamefont
  {Wang}}, \bibinfo {author} {\bibfnamefont {Y.~S.}\ \bibnamefont {Kivshar}}, \
  and\ \bibinfo {author} {\bibfnamefont {B.-Y.}\ \bibnamefont {Gu}},\ }\href
  {\doibase 10.1103/PhysRevLett.93.073901} {\bibfield  {journal} {\bibinfo
  {journal} {Phys. Rev. Lett.}\ }\textbf {\bibinfo {volume} {93}},\ \bibinfo
  {pages} {073901} (\bibinfo {year} {2004})}\BibitemShut {NoStop}%
\bibitem [{\citenamefont {Paulus}\ \emph {et~al.}(2000)\citenamefont {Paulus},
  \citenamefont {Gay-Balmaz},\ and\ \citenamefont
  {Martin}}]{Paulus-PRE-62-5797}%
  \BibitemOpen
  \bibfield  {author} {\bibinfo {author} {\bibfnamefont {M.}~\bibnamefont
  {Paulus}}, \bibinfo {author} {\bibfnamefont {P.}~\bibnamefont {Gay-Balmaz}},
  \ and\ \bibinfo {author} {\bibfnamefont {O.~J.~F.}\ \bibnamefont {Martin}},\
  }\href {\doibase 10.1103/PhysRevE.62.5797} {\bibfield  {journal} {\bibinfo
  {journal} {Phys. Rev. E.}\ }\textbf {\bibinfo {volume} {62}},\ \bibinfo
  {pages} {5797} (\bibinfo {year} {2000})}\BibitemShut {NoStop}%
\bibitem [{\citenamefont {Koenderink}(2010)}]{Koenderink:10}%
  \BibitemOpen
  \bibfield  {author} {\bibinfo {author} {\bibfnamefont {A.~F.}\ \bibnamefont
  {Koenderink}},\ }\href {\doibase 10.1364/OL.35.004208} {\bibfield  {journal}
  {\bibinfo  {journal} {Opt. Lett.}\ }\textbf {\bibinfo {volume} {35}},\
  \bibinfo {pages} {4208} (\bibinfo {year} {2010})}\BibitemShut {NoStop}%
\bibitem [{\citenamefont {Benson}(2011)}]{Benson2011}%
  \BibitemOpen
  \bibfield  {author} {\bibinfo {author} {\bibfnamefont {O.}~\bibnamefont
  {Benson}},\ }\href {\doibase 10.1038/nature10610} {\bibfield  {journal}
  {\bibinfo  {journal} {Nature}\ }\textbf {\bibinfo {volume} {480}},\ \bibinfo
  {pages} {193} (\bibinfo {year} {2011})}\BibitemShut {NoStop}%
\bibitem [{\citenamefont {Grigorenko}\ \emph {et~al.}(2008)\citenamefont
  {Grigorenko}, \citenamefont {Roberts}, \citenamefont {Dickinson},\ and\
  \citenamefont {Zhang}}]{Grigorenko-NATP-2-365W}%
  \BibitemOpen
  \bibfield  {author} {\bibinfo {author} {\bibfnamefont {A.~N.}\ \bibnamefont
  {Grigorenko}}, \bibinfo {author} {\bibfnamefont {N.~W.}\ \bibnamefont
  {Roberts}}, \bibinfo {author} {\bibfnamefont {M.~R.}\ \bibnamefont
  {Dickinson}}, \ and\ \bibinfo {author} {\bibfnamefont {Y.}~\bibnamefont
  {Zhang}},\ }\href {\doibase 10.1038/nphoton.2008.78} {\bibfield  {journal}
  {\bibinfo  {journal} {Nat. Phot.}\ }\textbf {\bibinfo {volume} {2}},\
  \bibinfo {pages} {365} (\bibinfo {year} {2008})}\BibitemShut {NoStop}%
\bibitem [{\citenamefont {Chang}\ \emph {et~al.}(2009)\citenamefont {Chang},
  \citenamefont {Thompson}, \citenamefont {Park}, \citenamefont {Vuleti\'{c}},
  \citenamefont {Zibrov}, \citenamefont {Zoller},\ and\ \citenamefont
  {Lukin}}]{chang:123004}%
  \BibitemOpen
  \bibfield  {author} {\bibinfo {author} {\bibfnamefont {D.~E.}\ \bibnamefont
  {Chang}}, \bibinfo {author} {\bibfnamefont {J.~D.}\ \bibnamefont {Thompson}},
  \bibinfo {author} {\bibfnamefont {H.}~\bibnamefont {Park}}, \bibinfo {author}
  {\bibfnamefont {V.}~\bibnamefont {Vuleti\'{c}}}, \bibinfo {author}
  {\bibfnamefont {A.~S.}\ \bibnamefont {Zibrov}}, \bibinfo {author}
  {\bibfnamefont {P.}~\bibnamefont {Zoller}}, \ and\ \bibinfo {author}
  {\bibfnamefont {M.~D.}\ \bibnamefont {Lukin}},\ }\href {\doibase
  10.1103/PhysRevLett.103.123004} {\bibfield  {journal} {\bibinfo  {journal}
  {Physical Review Letters}\ }\textbf {\bibinfo {volume} {103}},\ \bibinfo
  {eid} {123004} (\bibinfo {year} {2009})}\BibitemShut {NoStop}%
\bibitem [{\citenamefont {Righini}\ \emph {et~al.}(2007)\citenamefont
  {Righini}, \citenamefont {Zelenina}, \citenamefont {Girard},\ and\
  \citenamefont {Quidant}}]{Righini-Nat.Phys.-3-477}%
  \BibitemOpen
  \bibfield  {author} {\bibinfo {author} {\bibfnamefont {M.}~\bibnamefont
  {Righini}}, \bibinfo {author} {\bibfnamefont {A.~S.}\ \bibnamefont
  {Zelenina}}, \bibinfo {author} {\bibfnamefont {C.}~\bibnamefont {Girard}}, \
  and\ \bibinfo {author} {\bibfnamefont {R.}~\bibnamefont {Quidant}},\ }\href
  {\doibase 10.1038/nphys624} {\bibfield  {journal} {\bibinfo  {journal} {Nat.
  Phys.}\ }\textbf {\bibinfo {volume} {3}},\ \bibinfo {pages} {477} (\bibinfo
  {year} {2007})}\BibitemShut {NoStop}%
\bibitem [{\citenamefont {Barth}\ \emph {et~al.}(2010)\citenamefont {Barth},
  \citenamefont {Schietinger}, \citenamefont {Fischer}, \citenamefont {Becker},
  \citenamefont {N{\" u}sse}, \citenamefont {Aichele}, \citenamefont {L{\"
  o}chel}, \citenamefont {S{\" o}nnichsen},\ and\ \citenamefont
  {Benson}}]{Barth-Nano.Lett.-10-891}%
  \BibitemOpen
  \bibfield  {author} {\bibinfo {author} {\bibfnamefont {M.}~\bibnamefont
  {Barth}}, \bibinfo {author} {\bibfnamefont {S.}~\bibnamefont {Schietinger}},
  \bibinfo {author} {\bibfnamefont {S.}~\bibnamefont {Fischer}}, \bibinfo
  {author} {\bibfnamefont {J.}~\bibnamefont {Becker}}, \bibinfo {author}
  {\bibfnamefont {N.}~\bibnamefont {N{\" u}sse}}, \bibinfo {author}
  {\bibfnamefont {T.}~\bibnamefont {Aichele}}, \bibinfo {author} {\bibfnamefont
  {B.}~\bibnamefont {L{\" o}chel}}, \bibinfo {author} {\bibfnamefont
  {C.}~\bibnamefont {S{\" o}nnichsen}}, \ and\ \bibinfo {author} {\bibfnamefont
  {O.}~\bibnamefont {Benson}},\ }\href {\doibase 10.1021/nl903555u} {\bibfield
  {journal} {\bibinfo  {journal} {Nano. Lett.}\ }\textbf {\bibinfo {volume}
  {10}},\ \bibinfo {pages} {891} (\bibinfo {year} {2010})}\BibitemShut
  {NoStop}%
\bibitem [{\citenamefont {Canneson}\ \emph {et~al.}(2011)\citenamefont
  {Canneson}, \citenamefont {Mallek-Zouari}, \citenamefont {Buil},
  \citenamefont {Qu\'elin}, \citenamefont {Javaux}, \citenamefont {Mahler},
  \citenamefont {Dubertret},\ and\ \citenamefont
  {Hermier}}]{PhysRevB.84.245423}%
  \BibitemOpen
  \bibfield  {author} {\bibinfo {author} {\bibfnamefont {D.}~\bibnamefont
  {Canneson}}, \bibinfo {author} {\bibfnamefont {I.}~\bibnamefont
  {Mallek-Zouari}}, \bibinfo {author} {\bibfnamefont {S.}~\bibnamefont {Buil}},
  \bibinfo {author} {\bibfnamefont {X.}~\bibnamefont {Qu\'elin}}, \bibinfo
  {author} {\bibfnamefont {C.}~\bibnamefont {Javaux}}, \bibinfo {author}
  {\bibfnamefont {B.}~\bibnamefont {Mahler}}, \bibinfo {author} {\bibfnamefont
  {B.}~\bibnamefont {Dubertret}}, \ and\ \bibinfo {author} {\bibfnamefont
  {J.-P.}\ \bibnamefont {Hermier}},\ }\href {\doibase
  10.1103/PhysRevB.84.245423} {\bibfield  {journal} {\bibinfo  {journal} {Phys.
  Rev. B}\ }\textbf {\bibinfo {volume} {84}},\ \bibinfo {pages} {245423}
  (\bibinfo {year} {2011})}\BibitemShut {NoStop}%
\end{thebibliography}

%

\end{document}